 \documentclass[manuscript]{aastex}

\usepackage{multirow,natbib} %,epstopdf}
\slugcomment{Not to appear in Nonlearned J., 45.}

\shorttitle{{The Buried Starburst in II~Zw~096}}
\shortauthors{Inami et al.}

\begin{document}

\title{The Buried Starburst in the Interacting Galaxy \\
	II~Zw~096 as Revealed by the {\it Spitzer} Space Telescope}


% \author{Hanae Inami\altaffilmark{1,2,3} et al.}

\author{
H. Inami \altaffilmark{1,2,3},
L. Armus \altaffilmark{1}, 
J.A. Surace \altaffilmark{1},
J.M. Mazzarella \altaffilmark{4}, 
A.S. Evans \altaffilmark{5}, 
D.B. Sanders \altaffilmark{6},
J.H. Howell \altaffilmark{1,4}, 
A. Petric \altaffilmark{1},
T. Vavilkin \altaffilmark{7},
K. Iwasawa \altaffilmark{8}, 
S. Haan \altaffilmark{1}, 
E.J. Murphy \altaffilmark{1},
S. Stierwalt \altaffilmark{1},
P.N. Appleton \altaffilmark{9}, 
J.E. Barnes \altaffilmark{6}, 
G. Bothun \altaffilmark{10}, 
C.R. Bridge \altaffilmark{11}, 
B. Chan \altaffilmark{4}, 
V. Charmandaris \altaffilmark{12,13}, 
D.T. Frayer \altaffilmark{9},
L.J. Kewley \altaffilmark{6}, 
D.C. Kim \altaffilmark{5}, 
S. Lord \altaffilmark{4}, 
B.F. Madore \altaffilmark{4,14}, 
J.A. Marshall \altaffilmark{16}, 
H. Matsuhara \altaffilmark{2}, 
J.E. Melbourne \altaffilmark{11}, 
J. Rich \altaffilmark{6}, 
B. Schulz \altaffilmark{9}, 
H.W.W. Spoon \altaffilmark{16},
E. Sturm \altaffilmark{17}, 
V. U \altaffilmark{6,19}, 
S. Veilleux \altaffilmark{18}, 
K. Xu \altaffilmark{9}
}

\email{inami@ipac.caltech.edu}


\altaffiltext{1}{Spitzer Science Center, 
  California Institute of Technology, MS 220-6, Pasadena, CA 91125}
\altaffiltext{2}{Institute of Space and Astronautical Science(ISAS), 
  Japan Aerospace Exploration Agency(JAXA), Japan}
\altaffiltext{3}{Department of Space and Astronautical Science, 
  The Graduate University for Advanced Studies, Japan}
\altaffiltext{4}{Infrared Processing and Analysis Center, 
  California Institute of Technology, MS 100-22, Pasadena, CA 91125}
\altaffiltext{5}{National Radio Astronomy Observatory, 520 Edgemont Road, Charlottesville, VA 22903}
\altaffiltext{6}{Institute for Astronomy, University of Hawaii, 2680 Woodlawn Drive, Honolulu, HI 96822}
\altaffiltext{7}{Department of Physics and Astronomy, SUNY Stony Brook, Stony Brook, NY, 11794}
\altaffiltext{8}{ICREA and Universitat de Barcelona, Mart\'{i} i Franqu\`{e}s 1, Barcelona, Spain}
\altaffiltext{9}{NASA Herschel Science Center, California Institute of Technology, MS 100-22, Pasadena, CA 91125}
\altaffiltext{10}{Physics Department, University of Oregon, Eugene OR, 97402}
\altaffiltext{11}{California Institute of Technology, MS 320-47, Pasadena, CA 91125}
\altaffiltext{12}{Department of Physics, University of Crete, P.O. Box 2208, GR-71003, Heraklion, Greece}
\altaffiltext{13}{IESL/Foundation for Research and Technology - Hellas,  GR-71110, Heraklion, Greece and Chercheur Associ\'e, Observatoire de Paris, F-75014, Paris, France}
\altaffiltext{14}{The Observatories, Carnegie Institute of Washington, 813 Santa Barbara Street, Pasadena, CA 91101}
\altaffiltext{15}{The Jet Propulsion Laboratory, California Institute of Technology, Pasadena, CA 91125}
\altaffiltext{16}{Department of Astronomy, Cornell University, Ithaca, NY, 14953}
\altaffiltext{17}{MPE, Postfach 1312, 85741 Garching, Germany}
\altaffiltext{18}{Astronomy Department, University of Maryland, College Park, MD 20742}
\altaffiltext{19}{Harvard-Smithsonian Center for Astrophysics, 60 Garden St, Cambridge, MA 02138}

\begin{abstract}

An analysis of data from the {\it Spitzer} Space Telescope, {\it Hubble} Space Telescope,
{\it Chandra} X-ray Observatory, and {\it AKARI} Infrared Astronomy Satellite 
is presented for the $z=0.036$ merging galaxy system II~Zw~096 (CGCG~448-020).
Because II~Zw~096 has an infrared luminosity of log$(L_{IR}/L_\sun)=11.94$, 
it is classified as a Luminous Infrared Galaxy (LIRG), and was observed 
as part of the Great Observatories All-sky LIRG Survey (GOALS\footnote{http://goals.ipac.caltech.edu}). 
The {\it Spitzer} data suggest that 80\% of the total infrared luminosity 
%($ L_{IR(3-1100\micron)} = 2.84 \times 10^{38} \, \mathrm{W} = 7.42 \times 10^{11} \, L_\sun$) 
comes from an extremely compact, red source not associated with the nuclei of the 
merging galaxies.
The {\it Spitzer} mid-infrared spectra indicate no 
high-ionization lines from a buried active galactic nucleus in this source.
%although the $6.2 \, \micron$ polycyclic aromatic hydrocarbon (PAH) equivalent width 
%is low for a pure starburst nucleus. 
The strong detection of the $3.3 \, \micron$ and $6.2 \, \micron$ PAH emission features
in the {\it AKARI} and {\it Spitzer} spectra also implies that the energy source of II~Zw~096 is a starburst.
Based on {\it Spitzer} infrared imaging and {\it AKARI} near-infrared spectroscopy, 
the star formation rate is estimated to be $120 \, M_\sun \, \mathrm{yr^{-1}}$
and $> 45 \, M_\sun \, \mathrm{yr^{-1}}$, respectively.
Finally, the high-resolution {\it B}, {\it I}, and {\it H}-band images show many 
star clusters in the interacting system.
The colors of these clusters suggest at least two populations -- one with 
an age of $1-5$~Myr and one with an age of $20-500$~Myr, reddened by $0-2$ magnitudes 
of visual extinction. The masses of these clusters span a range between $10^{6}-10^{8} \, M_\sun$.
% The data suggest that II~Zw~096 may be an example of the type of buried, off-nuclear starburst first identified in 
% \object{NGC~4038/9} (\object{the Antennae Galaxies; Arp~244}), 
% but with a luminosity which is larger by more than an order of magnitude.
This starburst source is reminiscent of the extra-nuclear starburst seen in  
NGC~4038/9 (the Antennae Galaxies) and Arp~299 but 
approximately an order of magnitude more luminous than the Antennae. 
The source is remarkable in that the off-nuclear infrared luminosity dominates the enitre system.

\end{abstract}


\section{Introduction}

\object{II~Zw~096} (also known as \object{CGCG~448-020} or \object{IRAS~20550+1655})
%has an infrared luminosity $L_{IR}=L(8-1000\,\micron)=7.42 \times 10^{11} L_{\sun}$ and 
%a redshift of z=0.036.
has an infrared luminosity of log$(L_{IR}/L_{\sun})=11.94$ and 
a luminosity distance of 161~Mpc \citep{Armu09}. 
Since its luminosity
is above $L_{IR} \geq 10^{11} L_\sun$, it is classified as a
Luminous Infrared Galaxy, or LIRG. 
%A subsample of LIRGs, Ultra Luminous Infrared Galaxies (ULIRGs) define 
%the extreme of their population, having luminosities of $L_{IR} \geq 10^{12} L_\sun$. 
From its optical morphology, \object{II~Zw~096} appears to be a merger
of at least two gas-rich spirals.  
From the optical imaging, \cite{Arri04} classify \object{II~Zw~096} as 
a {\it Class~III} interacting galaxy, based on the system of \cite{Surace}, 
meaning the galaxy has two identifiable nuclei, with well developed tidal tails.

With near-infrared imaging and spectroscopy, 
\cite{Gold97} uncovered two extremely red sources to the east of the merging disks that 
appeared to mark the location of a highly obscured, young starburst.  These authors suggest
that \object{II~Zw~096} is one of a handful of LIRGs (such as \object{Arp~299} and \object{VV~114}) that are
experiencing enhanced star-formation before the final dissipative collapse of the system.
While \object{II~Zw~096} has previously been observed spectroscopically in the mid-infrared \citep{Dudl99},
the ground based data were not of high enough signal-to-noise to permit detection of PAH or the fine structure lines.
% \cite{Dudl99} has obtained $8-13\,\micron$ spectrum of \object{II~Zw~096} using 
% the Cooled Grating Spectrograph (CGS3) on the United Kingdom Infrared Telescope (UKIRT).
% However, due to the limited sensitivity, this author could not detect any features in this wavelength,
% such as the polycyclic aromatic hydrocarbon (PAH) feature at $11.3\,\micron$, [\ion{S}{4}], and [\ion{Ne}{2}].
It is also known that \object{II~Zw~096} hosts an OH megamaser at the B1950 position of
R.A.=20$\mathrm{^h}$55$\mathrm{^m}$05$\mathrm{^s}$.3, Dec=+16$\arcdeg$56$\arcmin$03$\arcsec$ 
\citep{Baan89}. \cite{Baan98} and \cite{Baan06} classified this megamaser as 
a starburst using the optical line ratios and the radio data. 

As part of the Great Observatories All-Sky LIRG Survey (GOALS), 
we have obtained images and spectra with the {\it Spitzer} Space Telescope, 
the {\it Hubble} Space Telescope ({\it HST}), the {\it Chandra} X-ray Observatory, 
and the {\it Galaxy Evolution Explorer} ({\it GALEX}) of a complete sample of 
LIRGs in the local universe. 
The GOALS targets consist of all LIRGs found in
the {\it IRAS} Revised Bright Galaxy Sample 
(RBGS; \cite{Sand03}), which covers galactic latitudes greater 
than five degrees and includes 629 extragalactic objects with $60\,\micron$ flux densities greater 
than 5.24~Jy. The median and the maximum redshift of the {\it IRAS} RBGS are z=0.008 and 0.088, respectively. 
The GOALS sample includes 179 LIRGs and 23 Ultra Luminous Infrared Galaxies (ULIRGs, $L_{IR} \geq 10^{12} L_\sun$), 
covering the full range of galaxy interaction stages from isolated spirals to late stage mergers.
GOALS provides an excellent dataset with which to explore 
the effect of mergers on infrared activity at low redshift.
A critical part of the GOALS survey has been to use the {\it Spitzer} Space 
Telescope to identify the location of the infrared emission within LIRGs, 
and to characterize the source of the power.
The GOALS project is fully described in \citet{Armu09}.

Here, we present the first mid and far-infrared imaging and spectroscopy of 
\object{II~Zw~096} from {\it Spitzer}, together with an analysis of 
the far-ultraviolet and optical imaging ({\it HST}), X-ray imaging ({\it Chandra}), 
and near-infrared spectra ({\it AKARI}).
In \S\ref{sec:obs}, we describe the observations and data reduction,
in \S\ref{sec:result},
we present our results, identifying the location of 
the bulk of the far-infrared emission, and in \S\ref{sec:dis} 
we discuss the nature of this buried power source.  
Cosmological parameters $H_0 = 70 \, \mathrm{km \, s^{-1} \,Mpc^{-1}}$, 
$\Omega_m = 0.28$, and $\Omega_\Lambda = 0.72$ are used throughout this paper.

\section{Observations and Data Reduction}\label{sec:obs}

\subsection{ {\it Spitzer} Imaging} \label{Spitzer}

{\it Spitzer} observations of \object{II~Zw~096} were taken with 
the Infrared Array Camera (IRAC; \cite{Fazi04}) at $3.6\, \micron$, $4.5\, \micron$, 
$5.8\, \micron$, and $8.0\, \micron$ on 2004 October 29
and the Multiband Imaging Photometer for 
Spitzer (MIPS; \cite{Riek04}) at $24\, \micron$, $70\, \micron$, and $160\, \micron$ 
on 2004 November 30 (PID~3672, PI~J.~Mazzarella).
All IRAC data were collected with 5 frames of $30 \sec$ high dynamic range (HDR) mode.
The $8.0\,\micron$ data saturated the $30 \sec$ integrations, so only the short $1.2 \sec$
frames were used. 
Total exposure times were $150 \sec$ at $3.6\, \micron$, $4.5\, \micron$,
and $5.8\, \micron$ and $6 \sec$ at $8.0\, \micron$.
The MIPS images were taken in photometry mode, using super-resolution at 
$70\, \micron$. Sixteen images of three seconds each were obtained.
The total exposure times were $48.2 \sec$, $37.7 \sec$, and $25.2 \sec$ for 
$24\, \micron$, $70\, \micron$, and $160\, \micron$, respectively.
A summary of the {\it Spitzer} observations is given in Table~\ref{tbl:obs}.

We used the MOPEX\footnote{Available from the {\it Spitzer} Science Center - 
see http://ssc.spitzer.caltech.edu/postbcd/mopex.html}
software package to process the IRAC and the MIPS images,
correcting the background, aligning, resampling the images, removing bad pixels,
combining into mosaics, and clipping cosmic-ray events.
Final mosaics were constructed with a pixel size of $0.6 \arcsec$/pixel for 
the IRAC images, $1.8 \arcsec$/pixel at $24\, \micron$,  
$4.0 \arcsec$/pixel at $70\, \micron$, and $8.0 \arcsec$/pixel at 
$160\, \micron$ for the MIPS images.

The WCS of the images are limited to $\sim 0.4 \arcsec$ by the telescope pointing accuracy.
However, alignment of the IRAC images has been refined 
with the Two Micron All Sky Survey (2MASS) to achieve a final accuracy ($\leq 0.3 \arcsec$).
For the MIPS images, the positional error is approximately 
$0.7 \arcsec$ based on the startracker-to-boresight uncertainty and 
the MIPS scan mirror uncertainty (Lacy M. et al. 2009, in preparation). 
A pointing refinement with 2MASS cannot be done 
because of the large wavelength difference between the MIPS and 2MASS bands.

All photometric fluxes were derived using the IDL APER routine.
For the IRAC images, we used $1.4\arcsec$ and $2.9\arcsec$ radii circular apertures,
using the aperture corrections from \cite{Sura04}.
% and the aperture correction values of $1.4\arcsec$ aperture are
% 1.71, 1.76, 2.15, and 2.39, and of $2.9\arcsec$ aperture are
% 1.15, 1.15, 1.25, and 1.43 for the
% $3.6\, \micron$, $4.5\, \micron$, $5.8\, \micron$, and 
% $8.0\, \micron$ image, respectively.}
For the MIPS images, we fit the PSF to the peak of emission to estimate total flux.

% The IRAC field of view (FOV) is 5.2 \arcmin $\times$ 5.2 \arcmin for all channels, 
% and the MIPS 24 \micron\ FOV is 5.4 \arcmin $\times$ 5.4 \arcmin,
% $70 \micron$ FOV is 2.6 \arcmin $\times$ 1.3 \arcmin, and 
% $160 \micron$ FOV is 0.53 \arcmin $\times$ 0.53 \arcmin.
% The pixel scale of IRAC mosaic images is 0.6\arcsec/pixel, and of 
% each MIPS mosaic images are 1.8\arcsec/pixel, 4.0\arcsec/pixel and 
% 8.0\arcsec/pixel, respectively.

\subsection{ {\it HST}} \label{HST}

The {\it HST} far-ultraviolet (FUV) and optical images were obtained with 
the Solar Blind Channel (SBC) and the Wide Field Camera (WFC) on 
the Advanced Camera for Surveys (ACS) on 2008 May 1 (PID~11196, PI~A.~Evans)
and 2006 April 15 (PID~10592, PI~A.~Evans), respectively.
The data were taken in the ACCUM mode using the LINE dither pattern for the WFC and 
the PARALLELOGRAM dither pattern for the SBC.
The near-infrared images were taken using the NIC2 camera on the Near Infrared 
Camera and Multi-Object Spectrometer (NICMOS) with the MULTIACCUM mode and 
the SPIRAL dither pattern on 2008 June 1 (PID~11235, PI~J.~Surace).
The following filters were used: 
ACS/SBC F140LP (FUV), ACS/WFC F435W ($B$-band), F814W ($I$-band), 
and NICMOS/NIC2 F160W ($H$-band) for 2528, 2520, 1440, and 2304 $\sec$, respectively.
We summarize the {\it HST} imaging observations in Table~\ref{tbl:obs}.

The {\it HST} data reduction began with the calibrated data products produced by STScI.
The ACS data were reduced with the PyDrizzle software 
included in IRAF/STSDAS provided by STScI, to identify and reject cosmic rays and bad pixels, 
to remove geometric distortion, and to combine the images into mosaics.
The NICMOS images were reduced using IRAF and IDL tasks designed to remove cosmic
rays and bad pixels, and correct for the different bias levels across the quadrants due to 
the four amplifiers.
The alignments of the {\it HST} images were improved based on 
the 2MASS astrometric solution. 
% First, we removed the cosmic-rays damaged pixels by the IRAF 
% cosmicrays package. Second, we used the text2mask to make a mask 
% for bad pixels and called fixpix to mask the bad pixels. Third, 
IRAF was used to construct the final mosaic, and the 
IDL routine APER was used to measure the photometry.
An aperture size of 0.25$\arcsec$, with corresponding aperture
corrections of 1.18, 1.16, 1.16, and 1.39, were used for photometric 
measurements of point sources in the final ACS/SBC F140LP, 
ACS/WFC F435W, F814W, and NICMOS F160W images, respectively.

\subsection{ {\it Chandra}} \label{Chandra}

X-ray observations of \object{II~Zw~096} were obtained with the AXAF CCD Imaging Spectrometer (ACIS) on 
the {\it Chandra} X-ray Observatory with VFAINT mode on 2007 September 10 (PID~8700551, PI~D.~Sanders).  
%We separated the image into the soft (0.5-2~keV) and the hard (2-7~keV) band X-ray images.
The background counts for total integration times $14.56 \, \mathrm{ks}$ are 
0.0025 counts/pixel and 0.004 counts/pixel in the soft ($0.5-2$~keV) and the hard ($2-7$~keV) bands, respectively.
The 5-$\sigma$ depths of the images are $1 \times 10^{-15} \, \mathrm{erg\,s^{-1}\,cm^{-2}}$ 
in the soft band and $5 \times 10^{-15} \, \mathrm{erg\,s^{-1}\,cm^{-2}}$ in the hard band.
%The pixel size of the images is 0.5$\arcsec$. %and the FWHM of the beam size is 0.5$\arcsec$. 
%The images have been smoothed using a Gaussian filter with 
%a $\sigma$ of 1.5 pixels (FWHM $\sim 1.8 \, \arcsec$).
%The integration time and observation date are given in Table~\ref{tbl:obs}.
The fundamental data reduction was done by the {\it Chandra} data analysis software CIAO version 3.4
and we have verified that there are no anomalous objects left in the data. 
Then we selected soft ($0.5-2$~keV) and hard ($2-7$~keV) X-ray energy ranges for 
making images and smoothed the data using
a Gaussian filter with a $\sigma$ of 1.5 pixels (FWHM $\sim 1.8 \arcsec$).

\subsection{ {\it Spitzer} IRS Spectra} \label{IRS}

\begin{figure}
  \begin{center}
    \includegraphics[scale=0.6, trim=50 250 0 0]{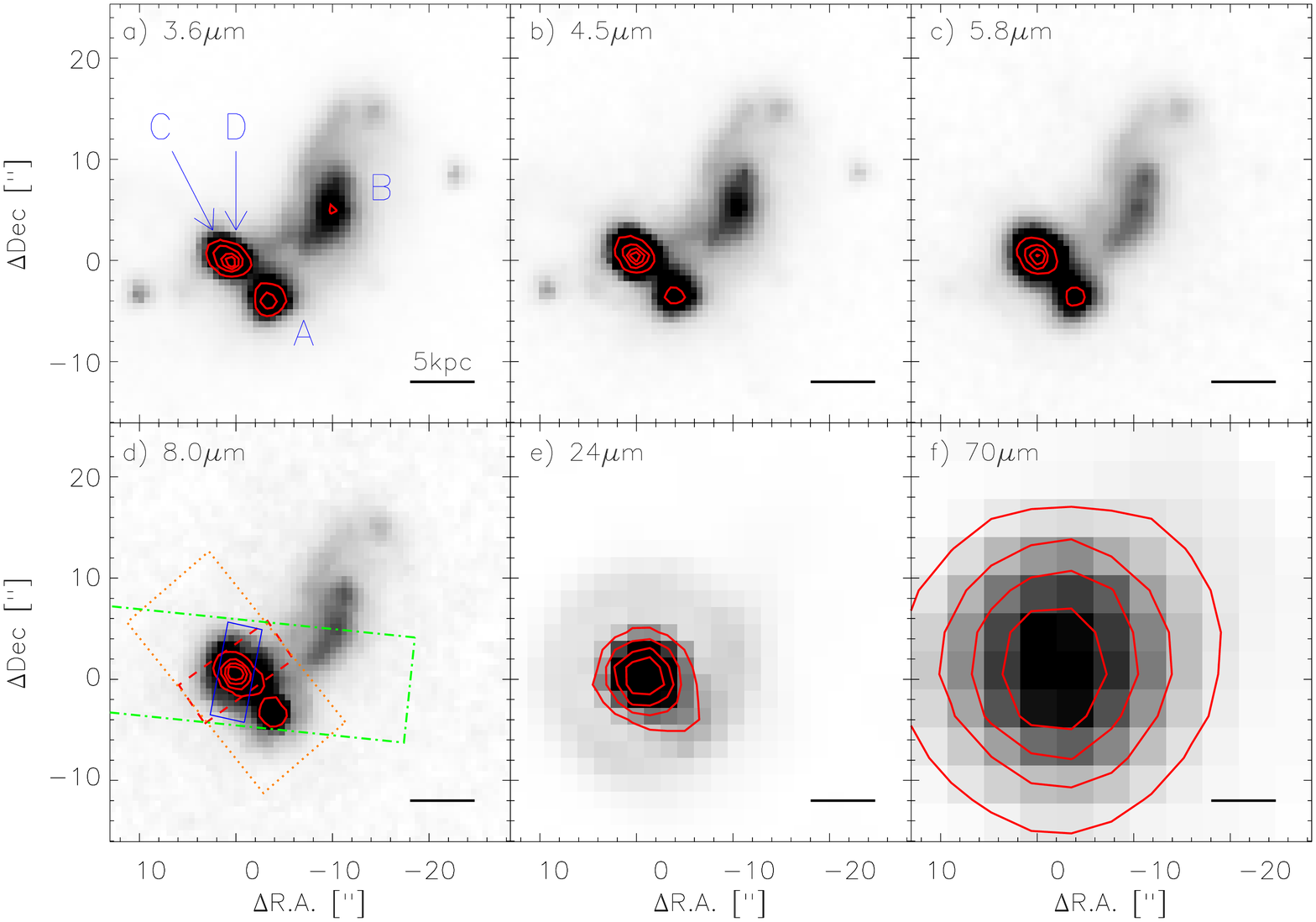}
    \caption{The {\it Spitzer} IRAC and MIPS images of II~Zw~096.  From upper
      left to lower right, these are (a) $3.6\, \micron$, (b) $4.5\, \micron$, (c) $5.8\, \micron$,
      (d) $8\, \micron$, (e) $24\, \micron$ and (f) $70\, \micron$. 
      In all frames, north is up and east is to the left.
      Sources A, B, C and D in II~Zw~096 are labeled in the $3.6\, \micron$ image, 
      using the same nomenclature as \cite{Gold97}.
      The origin of both axes in each panel corresponds to the position of source D at 
      R.A.=20$\mathrm{^h}$57$\mathrm{^m}$24$\mathrm{^s}$.34, Dec=+17$\arcdeg$07$\arcmin$39~1$\arcsec$ (J2000).
      The contour levels of the $3.6\, \micron$, $4.5\, \micron$, $5.8\, \micron$, 
      $8.0\, \micron$, $24\, \micron$ and $70\, \micron$ images start at 
      6.0, 9.0, 24, 55, 180, and 100 $\mathrm{MJy\,sr^{-1}}$, respectively, and increment with a step factor of two.
      The four IRS slits are overlayed on the IRAC $8\, \micron$ image in (d).
      The blue solid, red dashed, orange dotted and green dashed dotted boxes 
      indicate Short-Low, Short-High, Long-High and Long-Low, respectively
      (See the electronic edition of the Journal for a color version of this figure).
      Spectra extracted from the Short-Low and Short-High slits are from the peak emission only (source C+D), 
      not from the entire slit, as they are for the Long-Low and Long-High slits - see text for details.}
    \label{fig:spitzer_im}
  \end{center}
\end{figure}

\object{II~Zw~096} was observed 
with all modules of the {\it Spitzer} Infrared Spectrograph (IRS; \cite{Houc04}) 
in staring mode on 2006 November 12 (PID~30323, PI~L.~Armus). 
The slits were centered on the peak of the $8\,\micron$ emission.
The slit positions for all four modules are overlaid on the IRAC $8\, \micron$
image in Figure~\ref{fig:spitzer_im}.
The number of nod cycles and the integration time in each nod position 
are listed in Table~\ref{tbl:obs}.
% For the Short-Low spectrum ($5.2 - 14.5 \, \micron$, $R \sim 60 - 127$), 
% six cycles of ramp duration 14 $\sec$ has been coadded. 
% %This means that three cycles each of 14 $\sec$ in each nod position. 
% Coadded integration times of the other modules are four cycles and 14 $\sec$ for the Long-Low
% ($14.0 - 38.0 \, \micron$, $R \sim 60 - 127$), 
% six cycles and 30 $\sec$ for the Short-High 
% ($9.9 - 19.6 \, \micron$, $R \sim 600$), 
% and four cycles and 60 $\sec$ for the Long-High
% ($18.7 - 37.2 \, \micron$, $R \sim 600$) spectra.

The data were reduced using the S15.3 IRS pipeline at the {\it Spitzer} Science Center.
The pipeline software removes bad pixels and droop, subtracts the background, 
corrects linearity, and performs a wavelength and flux calibration. 
The backgrounds were subtracted with dedicated sky observations for the high-resolution data, 
and with off-source nods for the low-resolution data.
One dimensional spectra were extracted with the 
SPICE\footnote{http://ssc.spitzer.caltech.edu/postbcd/spice.html} software package, 
using the standard point source extraction apertures for all slits.  
For the high-res slits, this is equivalent to a full-slit extraction.
%from two-dimensional to one-dimensional data.
The slit widths of Short-Low (SL), Long-Low (LL), 
Short-High (SH), and Long-High (LH) are $3.6\arcsec$, $10.5\arcsec$, 
$4.7\arcsec$, and $11.1\arcsec$, respectively.
The extraction size for SL is $9.5\arcsec$, LL is $38.3\arcsec$, 
SH is $11.7\arcsec$, and LH is $22.8\arcsec$.

\subsection{ {\it AKARI} Near-Infrared Spectra} \label{AKARI}

The near-infrared spectra of \object{II~Zw~096} were 
taken with {\it AKARI} on 2009 May 12 and 13 (PID~GOALS, PI~H.~Inami)
during the post-helium (warm phase) mission using the Infrared Camera (IRC; \cite{Onak07}). 
The $2.5-5~\micron$ spectra were obtained using the $1\arcmin \times 1\arcmin$ 
aperture (the Np aperture) using a moderate-resolution grism.
The spectral resolving power is $\sim 120$ at $3.5\,\micron$ for 
a point-like source ($\sim 5\arcsec$ FWHM).
The use of the Np aperture is intended to avoid the confusion of spectra but 
multiple sources in this aperture may interfere each other (see \S\ref{subsec:AKARI_spec}).
The observations used the IRCZ4 Astronomical Observation Template,
taking four spectra, a reference image, and then five spectra at the end
of the sequence. The exposure time for each spectrum was 44.41 seconds.
Two pointings were used (in total 2 pointings $\times$ 9 frames) for 
the observations and the total exposure time was 799 seconds.

The data reduction was performed with the standard software package provided by 
JAXA, the IRC Spectroscopy Toolkit for Phase 3 data Version 
20090211\footnote{This is available from http://www.ir.isas.ac.jp/ASTRO-F/Observation/DataReduction/IRC/}.
This software performs dark subtraction, linearity correction, flat correction, 
background subtraction, wavelength calibration, spectral inclination correction 
and spectral response calibration.
We extracted the 1D spectra from the 2D spectral image 
using an aperture three pixels wide ($4.4\arcsec$).
The number of bad pixels is increasing dramatically in the post-helium mission
due to the elevated detector temperatures.
A $3 \times 3$  moving window was used to isolate and replace bad pixels
in each spectral image before co-adding the data and extracting the 1D spectrum.
The wavelength accuracy is evaluated to be $0.01\,\micron$ and the
uncertainty in absolute flux calibration is estimated to be
$\sim$10\% \citep{Ohya07}.

\section{Results}\label{sec:result}

\subsection{Spitzer Imaging} \label{IRemission}

% Using the {\it Spitzer} Space Telescope, we have been obtained the first mid- and 
% far-infrared images of \object{II~Zw~096}.
In Figure~\ref{fig:spitzer_im}, the {\it Spitzer} images from 
IRAC $3.6\, \micron$ through MIPS $70\, \micron$ are presented
in order of increasing wavelength.
Three distinct objects are seen in the IRAC data. 
The nuclei of the merging galaxies are to the north (source B) and south (source A) \citep{Gold97}.
In addition there is a bright, red source to the east (sources C and D), 
which is resolved and elongated at an angle of roughly $60^{o}$.
This resolved IRAC source is coincident with the near-infrared 
sources ``C" (R.A.=20$\mathrm{^h}$57$\mathrm{^m}$24$\mathrm{^s}$.47, Dec=+17$\arcdeg$07$\arcmin$39~9$\arcsec$) and 
``D'' (R.A.=20$\mathrm{^h}$57$\mathrm{^m}$24$\mathrm{^s}$.34, Dec=+17$\arcdeg$07$\arcmin$39~1$\arcsec$)  from \cite{Gold97}.
%sources ``C" (R.A.=20:57:24.5, Dec=+17:07:39.8) and 
%``D'' (R.A.=20:57:24.3, Dec=+17:07:39.1) from \cite{Gold97}.
We estimate the flux density of source D 
at $3.6\, \micron$, $4.5\, \micron$, $5.8\, \micron$, and $8.0\, \micron$, as 
$2.85 \pm 0.08$, $4.67 \pm 0.13$, $12.7 \pm 0.2$, and $41.9 \pm 0.8$~mJy, respectively,
using a circular aperture size of $1.4\arcsec$ radius.

\begin{figure}
  \begin{center}
    \includegraphics[scale=0.5]{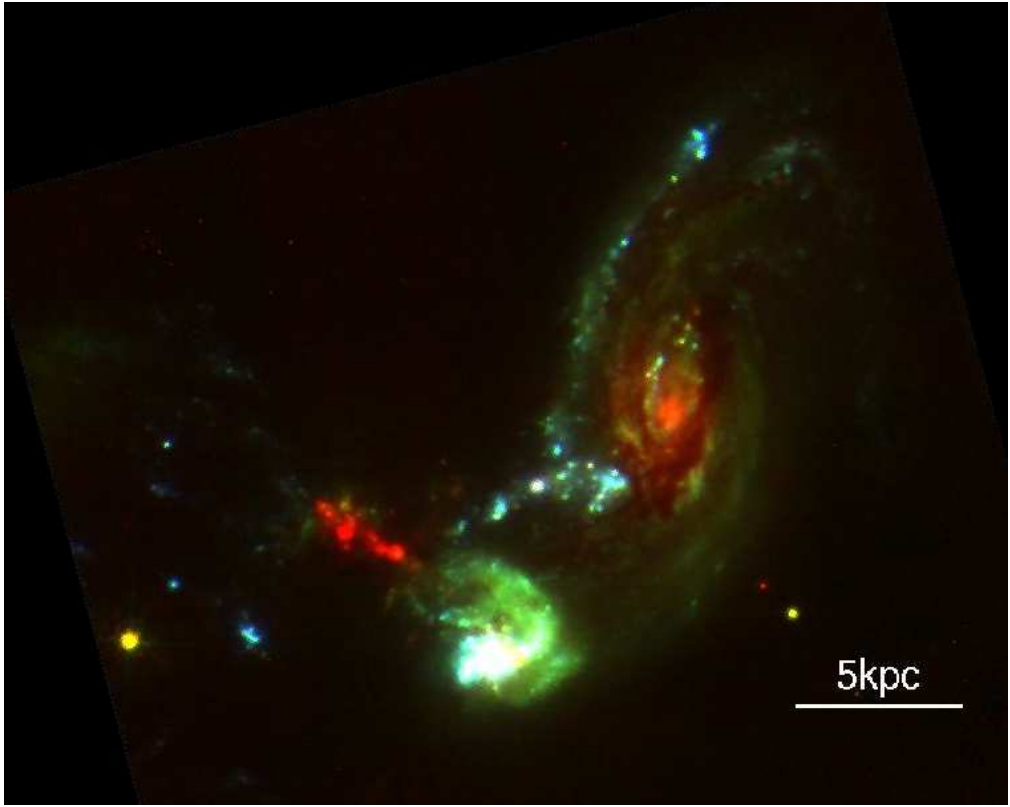}
    \caption{False-color RGB image of II~Zw~096 made by combining 
      the {\it HST} ACS SBC F140LP (blue), ACS F435W (green) and NICMOS F160W (red) images.}
    \label{fig:RGB_im}
  \end{center}
\end{figure}

The MIPS $24\, \micron$ image shows a strong point source coincident with 
the position of source D, along with a lot of emission to the southwest, 
coincident with source A.
Although the MIPS $70\, \micron$ image is unresolved, 
the position of the peak is consistent with source D.
PSF fitting to source D at $24\,\micron$ and $70\,\micron$ suggests that 
it produces $67.2 \pm 10.6 \%$ ($1.37 \pm 0.18$~Jy) and 
$87.9 \pm 13.8 \%$ ($10.2 \pm 1.2$~Jy) of the total galaxy 
flux at $24\, \micron$ and $70\, \micron$, respectively.
The angular separation of sources C and D is $\sim2.0\arcsec$
(see Figure~\ref{fig:HST_im} or \ref{fig:RGB_im}).
Although this small separation cannot be resolved with MIPS,
the centroid of source D can be determined much more accurately 
than the FWHM of the beam.
In addition, we have simulated the MIPS image using two point sources
separated by $2.0\arcsec$, determining that a flux ratio of $4:1$ ($\mathrm{D}:\mathrm{C}$) is 
required to match the data.
The photometry is summarized in Table~\ref{tbl:phot}.
% In addition, we have done a simulation to check 
% what flux ratio of two $2.0\arcsec$ separated point sources is needed to generate similar contours 
% to the MIPS $24\,\micron$ image. This simulation gives the result that 
% these two sources should have at least a flux ratio of $4:1$.
% The results of the photometries are summarized in Table~\ref{tbl:phot}.

We estimate the total infrared (TIR, $3-1100 \, \micron$) luminosity
of source D to be $TIR = 5.85 \times 10^{11} \, L_\sun$ from the {\it Spitzer} 24 and $70\,\micron$ 
images, using the formula Eq.(4) in \cite{DH02}, assuming that the $160\, \micron$ flux of 
source D is 87.9\% of the total at $70\, \micron$. 
This is approximately 80\% of the TIR of the system.
\object{II~Zw~096} has an infrared luminosity (LIR, $8-1000\,\micron$) of 
$10^{11.94}L_\odot$ \citep{Armu09}.
Assuming source D accounts for $\sim80$\% of the total LIR of the system
(the same fraction as TIR), we estimate an LIR of $6.87\times10^{11}\,L_\sun$.

\subsection{ {\it HST} Imaging } \label{HSTimage}

\begin{figure}
  \begin{center}
    \includegraphics[scale=0.5]{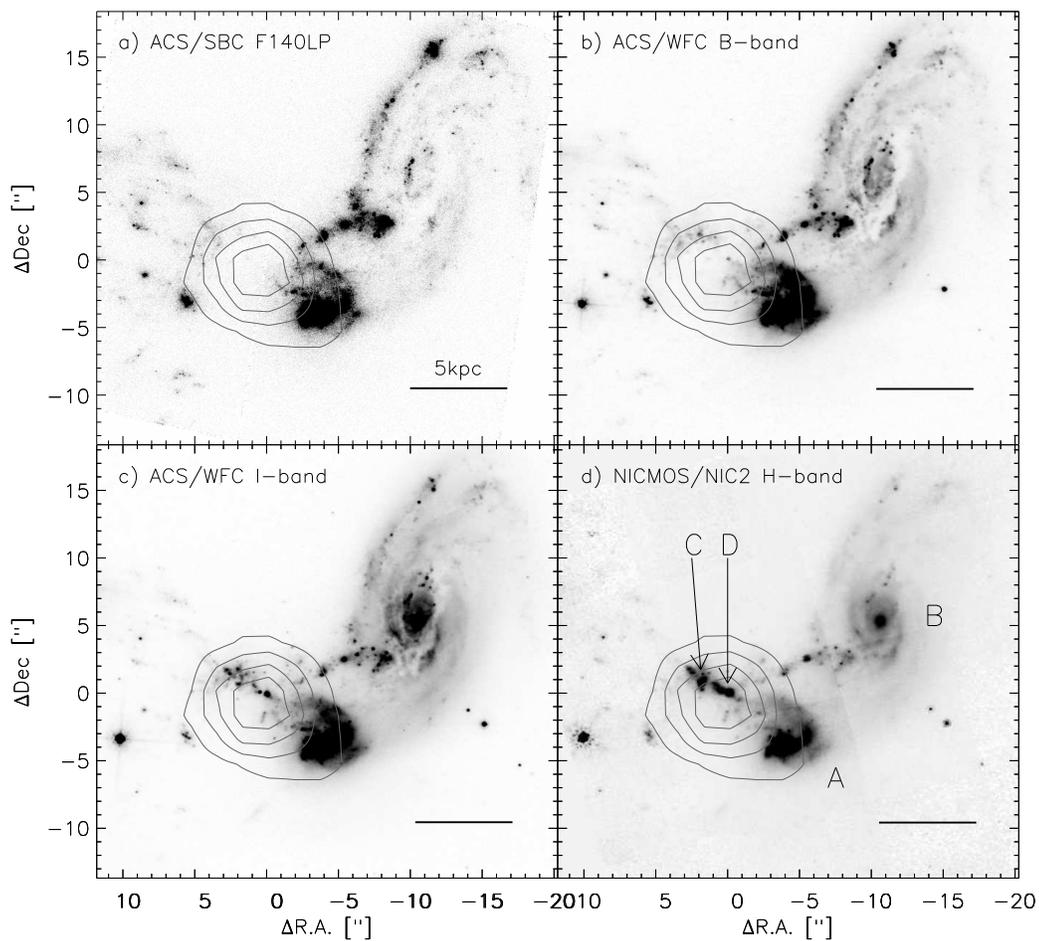}
    \caption{ The {\it HST}/ACS and SBC F140LP, F435W, F814W, and NICMOS F160W 
      grey-scale images of II~Zw~096 with {\it Spitzer} MIPS 24 micron contours overlaid. 
%       The size of the cross bar at the center 
%       of the contours in each image indicates the 1-2 $\sigma$ uncertainties 
%       of the MIPS pointing reconstruction, approximately 0.3 \arcsec.
      The source of the prodigious far-infrared emission is mostly invisible in the optical and UV.
      The labels $\Delta \mathrm{R.A.} = 0$ and $\Delta \mathrm{Dec} = 0$ indicate the coordinate of source D
      (R.A.=20$\mathrm{^h}$57$\mathrm{^m}$24$\mathrm{^s}$.34, Dec=+17$\arcdeg$07$\arcmin$39~1$\arcsec$).}
    \label{fig:HST_im}
  \end{center}
\end{figure}

{\it HST}/ACS and NICMOS images of \object{II~Zw~096} are shown in Figure~\ref{fig:HST_im}, 
from the upper left to the lower right, ACS/SBC F140LP (FUV), 
ACS/WFC F435W ($B$-band), F814W ($I$-band), and NICMOS/NIC2 F160W ($H$-band).
The contours of the {\it Spitzer} $24\, \micron$ image are overlaid in each panel.

The $I$-band and the $H$-band images with the $24\,\micron$ contour overlayed in Figure~\ref{fig:HST_im} 
suggest that source D is the origin of the bulk of the far-infrared emission in \object{II~Zw~096}.
The contour of the MIPS $70\, \micron$ image is also consistent with the position of source D.

\begin{figure}
  \begin{center}
    \includegraphics[scale=1.0, trim=20 30 150 20]{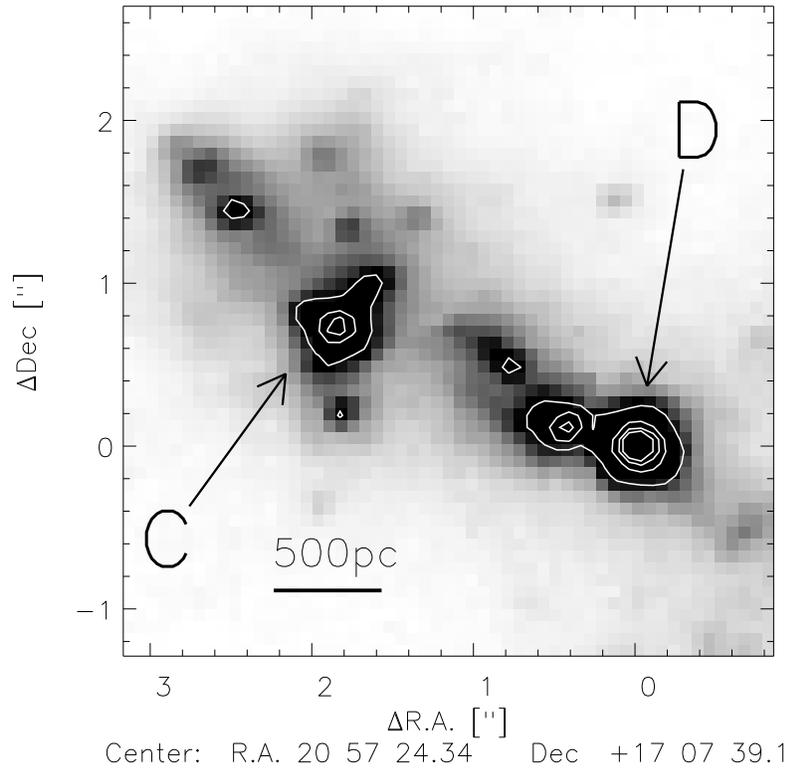}
    \caption{Close-up of the NICMOS image, identifying the red sources 
      designated as ``C'' and ``D'' by \cite{Gold97}.
      The new {\it HST} data reveal not only a large number of young, blue clusters, 
      but also the true, complex nature of the buried starburst.}
    \label{fig:zoomCD_im}
  \end{center}
\end{figure}

The {\it HST} NICMOS images resolve the structure of the red sources into at least 
10 ``knots'' (see Figure~\ref{fig:zoomCD_im}). Most of these knots
are not seen in the visible image due to heavy extinction.
In the $H$-band, the absolute magnitudes of sources C and D are 
$M_H = -18.7 \pm 0.1 \, \mathrm{mag}$ and $-19.5 \pm 0.1 \, \mathrm{mag}$, respectively,
measured with an $0.25\arcsec$ radius aperture.
We estimate that approximately $0.81\pm 0.04\%$ and $1.62\pm 0.06\%$ of 
the $H$-band flux of the entire system comes from sources C and D alone. 
Sources C and D provide about  $7.78 \pm 0.39\%$ and $15.5 \pm 0.6\%$ of the
$H$-band emission for the red complex in Figure~\ref{fig:zoomCD_im}, respectively.

%The peak of the infrared emission is completely invisible in the far-UV and 
%the $B$-band optical images. 
In Figure~\ref{fig:RGB_im}, we show a false-color RGB image of \object{II~Zw~096} made
from the NICMOS $H$-band, ACS $B$-band and ACS FUV images.
%In this image, the extremely red complex containing sources C and D is striking.
The extremely red sources C and D are obvious in this figure, as are a number 
of unresolved blue sources to the west.  These are presumably young star clusters in 
the northern spiral arm of the western galaxy and the overlap region.

\subsection{Chandra Imaging} \label{Chandra_image}

\begin{figure}
  \begin{center}
    \includegraphics[scale=0.5]{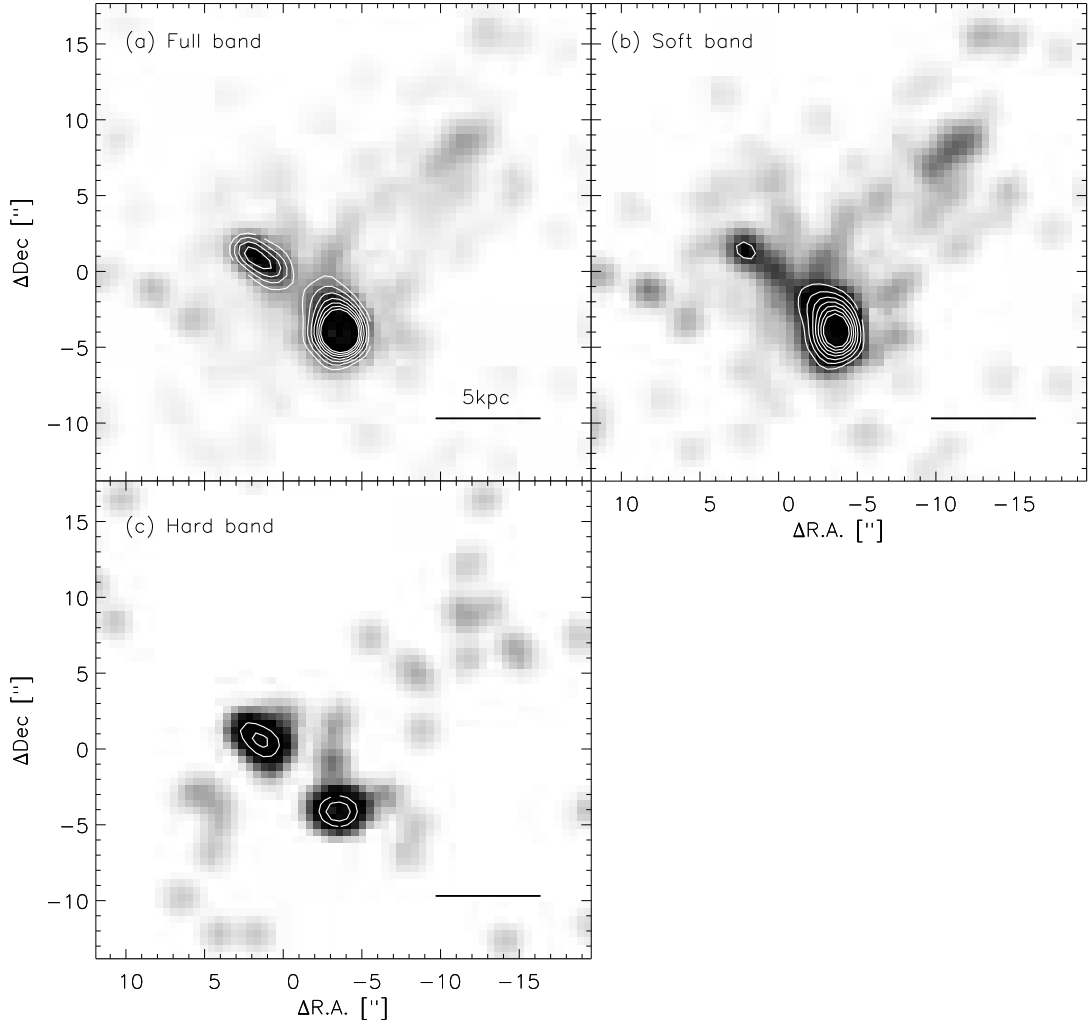}
    \caption{ The {\it Chandra} (a) full band ($0.5-7$~keV), (b) soft X-ray ($0.5-2$~keV), and 
      (c) hard X-ray ($2-7$~keV) images. 
%      The {\it Spitzer} MIPS 24 micron contour is overlaid. 
      The contour levels of all of the images start at 0.3 with a step of 0.2.
      The labels $\Delta \mathrm{R.A.} = 0$ and
      $\Delta \mathrm{Dec} = 0$ indicate the coordinate of source D
      (R.A.=20$\mathrm{^h}$57$\mathrm{^m}$24$\mathrm{^s}$.34, Dec=+17$\arcdeg$07$\arcmin$39~1$\arcsec$).}
    \label{fig:Chandra_im}
  \end{center}
\end{figure}

The {\it Chandra} X-ray images are shown in Figure~\ref{fig:Chandra_im}.
In the full ($0.5-7$~keV) and soft ($0.5-2$~keV) bands, 
source A dominates the emission.  In the hard-band ($2-7$~keV) image, 
sources C and D become much more prominent.
The $0.5-2$~keV luminosity of source A and source C+D complex are 
$6.6 \times 10^{40} \, \mathrm{erg \, s^{-1}}$ and 
$1.8 \times 10^{40} \, \mathrm{erg \, s^{-1}}$, respectively.  
In the hard band, the $2-7$~keV luminosity of 
source A is $1 \times 10^{41} \, \mathrm{erg \, s^{-1}}$ and
the total of sources C and D is $1 \times 10^{41} \, \mathrm{erg \, s^{-1}}$.

The C+D complex is clearly extended (see Figure~\ref{fig:Chandra_im}).
Source C dominates the soft-band emission from the complex,
with the fraction of the X-ray emission coming from source D 
apparently rising towards the hard band.  This is evidenced by 
a gradual shifting of the peak from NE to SW in Figure~\ref{fig:Chandra_im}.
However, given the low number of total counts ($\sim20$) from this region and 
the strong overlap of source C and D, it is difficult to derive an independent
measure of the hardness ratio of C and D, separately.
Together, the hardness ratio (Hard-Soft/Hard+Soft) of source C+D is $0.05 \pm 0.14$.

% The hardness ratios  (the difference of the hard and soft X-ray counts, 
% divided by the sum of the hard and soft-band counts; (Hard-Soft)/(Hard+Soft)) 
% of the two are $-0.04 \pm 0.19$ and $0.00 \pm 0.24$, respectively.}

%Source C is brighter with 27 counts than Source D with 20 counts. 
%Also Source C looks more peaky in the hard band. 
%However, these need to be verified by data with a longer exposure.

%the absorbing column density of the NE source is a few times 1e22 cm-2 in nH, 
%which is consistent with that the silicate absorption implies.

\subsection{ {\it Spitzer} IRS Spectroscopy} \label{spectra}

\begin{figure}[htbp]
  \begin{minipage}{0.5\hsize}
    \begin{center}
      \includegraphics[width=85mm]{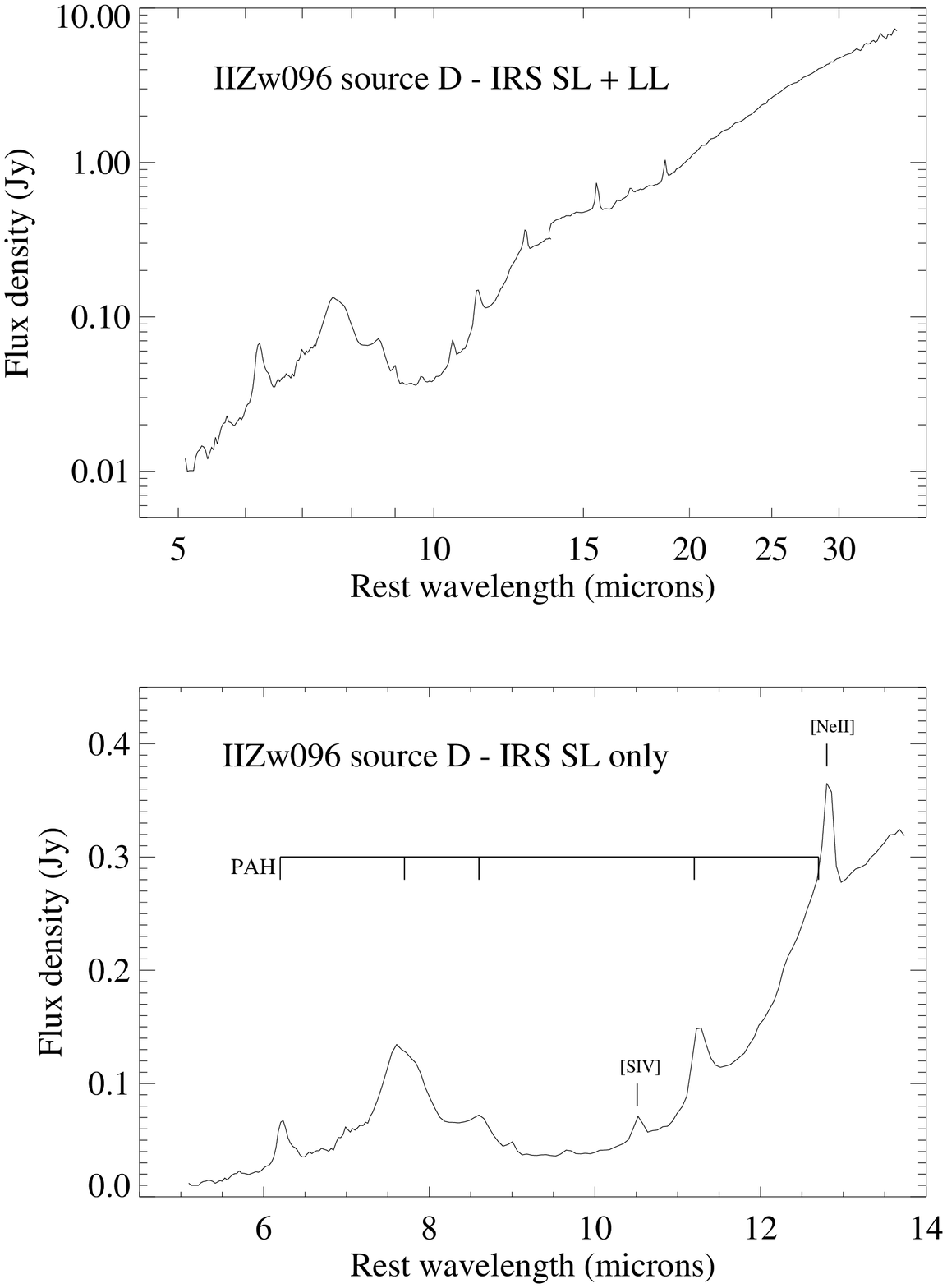}
    \end{center}
  \end{minipage}
  \begin{minipage}{0.5\hsize}
    \begin{center}
      \includegraphics[width=80mm]{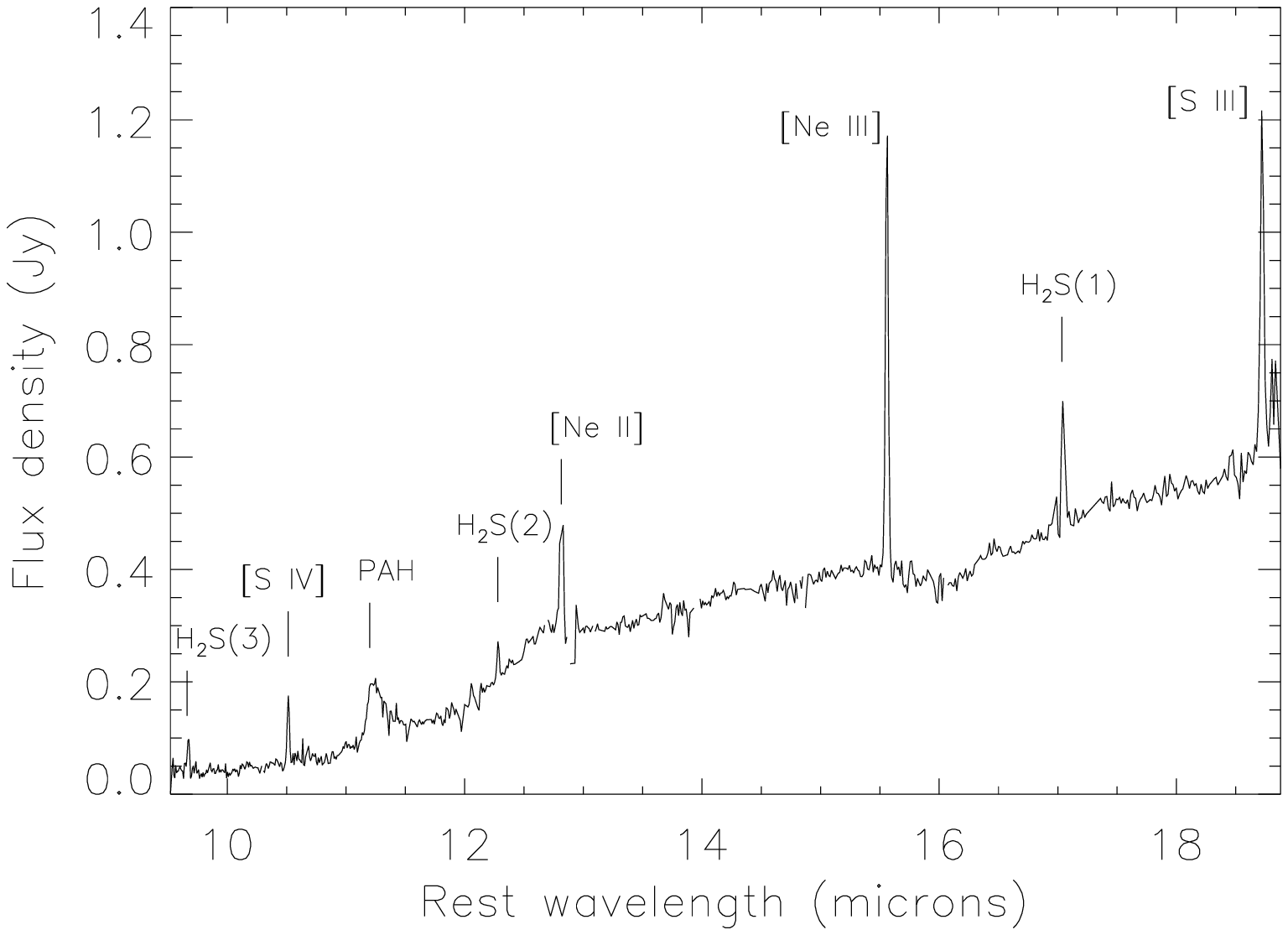}
      \includegraphics[width=80mm]{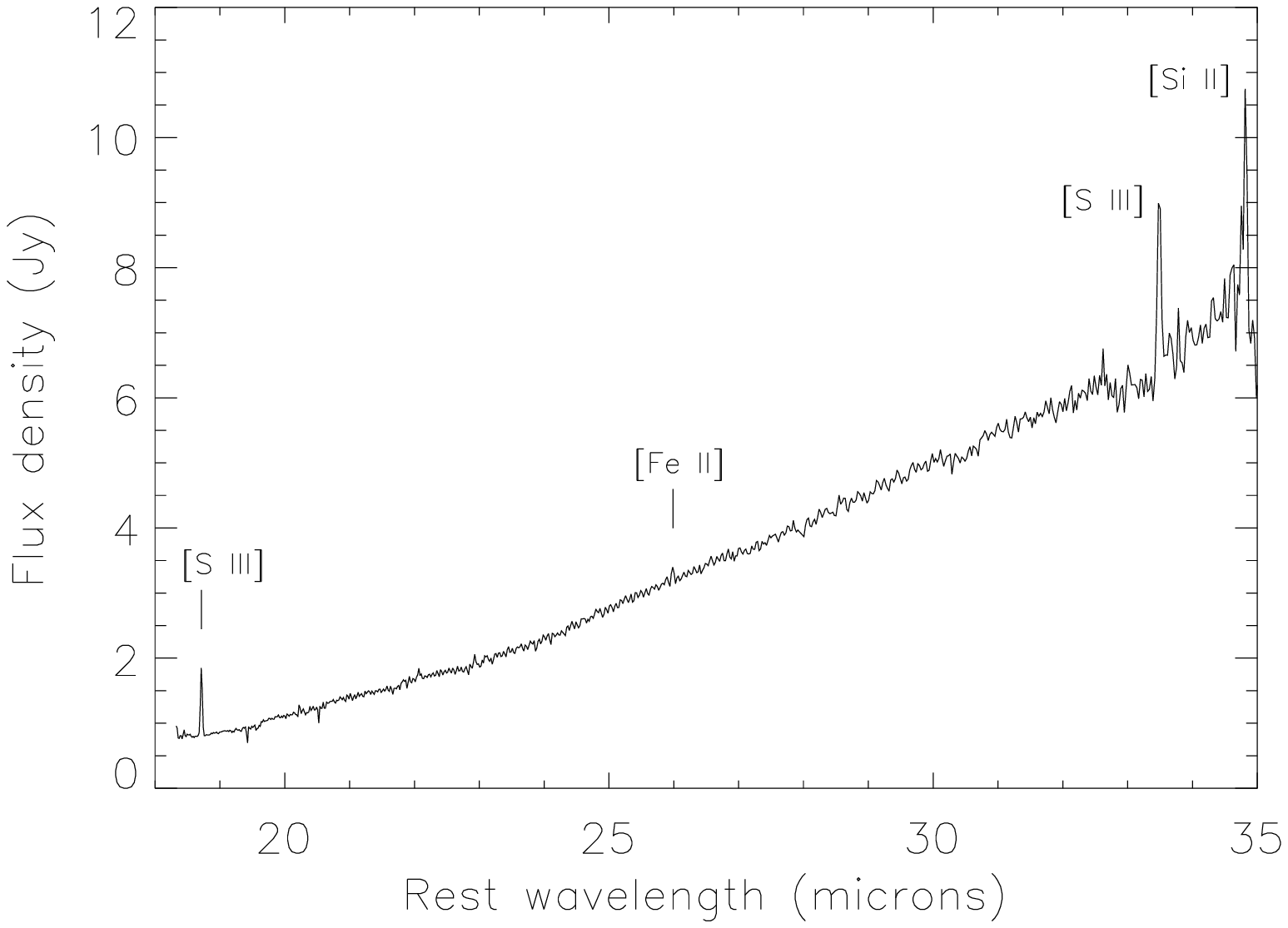}
    \end{center}
  \end{minipage}
    \caption{{\it Spitzer} IRS spectra of source D (see Fig~\ref{fig:zoomCD_im}).
      The full low-resolution spectrum is at the top left, 
      while a close-up of the Short-Low spectrum is shown at the bottom left.
      The Short-High and Long-High spectrum are shown at the top right and 
      the bottom right, respectively. Some residual order tilting is evident in the two reddest orders
      of Long-High.  All spectra are shown in the rest frame.}
    \label{fig:IRS}
\end{figure}

The {\it Spitzer} IRS Short-Low (SL), Long-Low (LL), 
Short-High (SH), and Long-High (LH) spectra of sources C and D (C+D) are 
shown in Figure~\ref{fig:IRS}, and the fluxes of the detected lines are shown in 
Table~\ref{tbl:spectra} and \ref{tbl:PAHFIT}. 
Note that the LL and LH spectra cover not only source C+D but also source A 
due to their limited resolutions (see figure~\ref{fig:spitzer_im}~(d)).
However, the scale factor of LL/SL is 1.09, indicating the continuum is dominated
by source C+D only.
The IRS low-resolution spectra are dominated by strong PAH emission and 
silicate absorption. The high-resolution data show strong fine structure line emissions
such as [\ion{Ne}{3}], [\ion{Ne}{2}], and [\ion{S}{3}].

%SPLINE RESULTS
%-------------

%Low-res results
From the SL data, we measure $6.2\, \micron$ and 
$11.2\, \micron$ PAH equivalent widths (EQW) of 0.26 and
0.14 $\micron$, respectively, and line fluxes of
$(5.57 \pm 0.19) \times 10^{-16} \, \mathrm{W \, m^{-2}}$ and
$(3.24 \pm 0.32) \times 10^{-16} \, \mathrm{W \, m^{-2}}$, respectively, using
a simple spline fit to the local continuum under the features.
The apparent silicate optical depth at $9.7\, \micron$
is $ \tau_{9.7} \sim 1 $ when a screen geometry is assumed.
This implies an $A_{V} \geq 19 ~\mathrm{mag}$ toward source C+D \citep{Roch84}.
This is comparable with the estimate
of $1-2$~mag in the $K$-band from the near-infrared colors \citep{Gold97}.

%High-res results
We do not see evidence for $14.3\, \micron$ [\ion{Ne}{5}] emission in the SH spectrum.
A 3-$\sigma$ upper limit of the [\ion{Ne}{5}] line flux is $< 0.10 \times 10^{-16} \, \mathrm{W\,m^{-2}}$.
The [\ion{Ne}{2}] line has a flux of $(2.74 \pm 0.47) \times 10^{-16} \, \mathrm{W\,m^{-2}}$, 
implying a 3-$\sigma$ upper limit on the [\ion{Ne}{5}]/[\ion{Ne}{2}] line flux ratio of 0.04.
The $25.9\, \micron$~[\ion{O}{4}] line is not detected, with a 3-$\sigma$ upper limit of
$< 0.29 \times 10^{-16} \, \mathrm{W\,m^{-2}}$, implying an upper limit on the [\ion{O}{4}]/[\ion{Ne}{2}]
line flux ratio of 0.11.
The $18.71\, \micron$ [\ion{S}{3}] line fluxes are detected in both of SH and LH: 
$(2.16 \pm 0.22) \times 10^{-16} \, \mathrm{W\,m^{-2}}$ in SH and  
$(3.27 \pm 0.04) \times 10^{-16} \, \mathrm{W\,m^{-2}}$ in LH.
The large value for $18.71\, \micron$ [\ion{S}{3}] in LH implies that 
there is measurable emission line flux from source A.
The flux of the [\ion{S}{3}] line at $33.48\, \micron$ is 
$(4.77 \pm 0.07) \times 10^{-16} \, \mathrm{W\,m^{-2}}$.
Using the [\ion{S}{3}] 18.71/[\ion{S}{3}] 33.48 line flux ratio in LH of 0.69 $\pm$ 0.01,
we derive an ionized gas density of  $250 \, \mathrm{cm^{-3}}$ at $10^4$~K. 
The density is an average value over the area enclosed by LH which includes sources C, D, and A.
% Since the LH aperture is so large and includes sources C, D, and A, the [\ion{O}{4}]/[\ion{Ne}{2}] and the
% [\ion{S}{3}] ratio are really averages over these regions (see Figure~\ref{fig:spitzer_im} (d)).
% The fact that the $24\,\micron$ emission is dominated by source C+D and 
% that the SL and LL spectra agree very well (to better than 10\%) at the overlap region 
% suggests that the dust continuum has very little contribution from source A.

The fluxes of H$_2~S(3)$, H$_2~S(2)$, and H$_2~S(1)$ lines are 
$(0.45 \pm 0.05) \times 10^{-16}$ W m$^{-2}$, 
$(0.29 \pm 0.03) \times 10^{-16}$ W m$^{-2}$, and 
$(0.73 \pm 0.05) \times 10^{-16}$ W m$^{-2}$, respectively.
Taking these lines, and assuming an ortho to para ratio of 3.0, we
estimate the warm molecular gas has a temperature of 329 K, and a
total mass of $4.5 \times 10^{7} \, M_\sun$.

% PAHFIT RESULTS
%----------------
%In addition of this measurement from the spline method, we also use 
%PAHFIT (\cite{Smit07}) to obtain 6.2$\micron$, 7.7$\micron$ complex, 
%8.3$\micron$, 8.6$\micron$, 11.3$\micron$ complex, 12.6$\micron$, 
%13.6$\micron$, 14.2$\micron$ and 17$\micron$ complex PAH fluxes and EQWs.
%Because the PAHFIT measurement includes the line wings and corrections for 
%reddening, the values should be larger than the spline measurements.
%The results are given in Table~\ref{tbl:PAHFIT}.

If we fit the entire IRS low-res spectrum using PAHFIT (\cite{Smit07}), we can estimate
a total PAH flux (see Table~\ref{tbl:PAHFIT}) and compare this to the far-infrared flux which 
we estimate originates from source C+D.
When this is done, we derive a ratio of $F_{PAH_{tot}}/F_{IR} \sim 0.02$. 
Since the continuum emission at the overlap wavelengths
of SL and LL have a difference less than 10\%, 
dust emission in our low-res spectrum is dominated by source C+D.

\subsection{ {\it AKARI} Near-Infrared Spectroscopy} \label{subsec:AKARI_spec}

\begin{figure}[htbp]
  \begin{center}
    \includegraphics[scale=0.9]{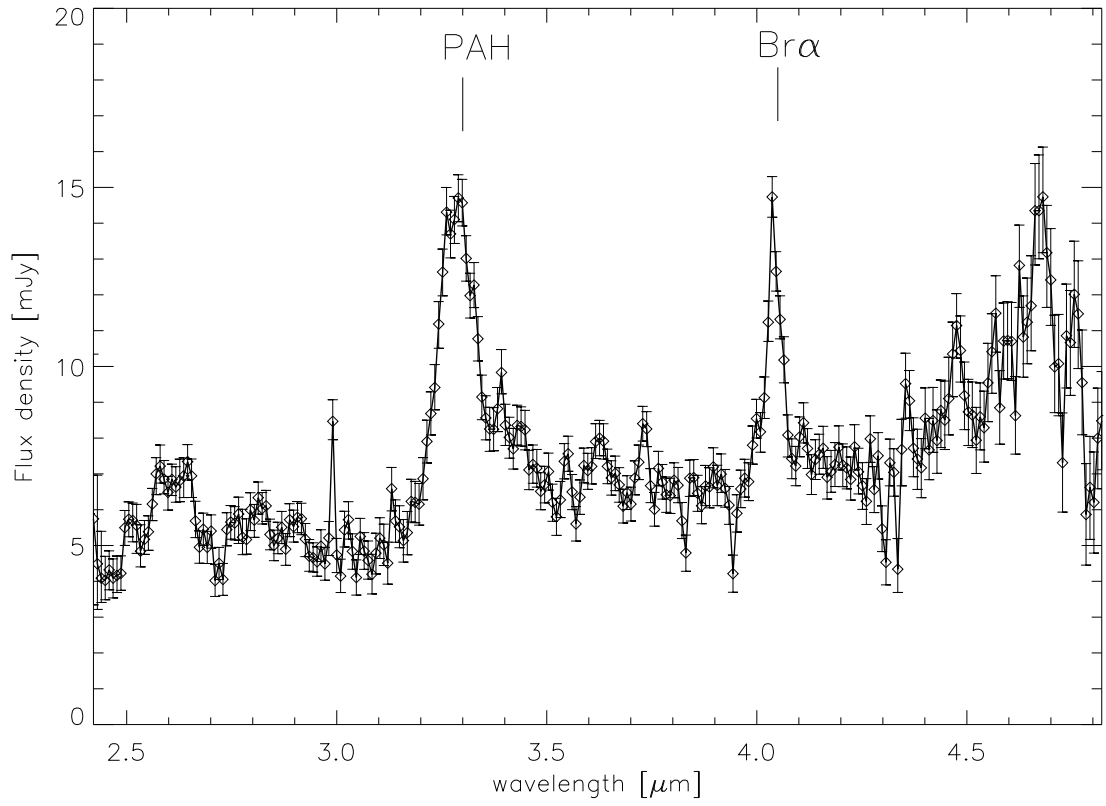}
    \caption{{\it AKARI} near-infrared spectrum of sources C and D.
      The wavelength scale is given in the rest-frame.}
    \label{fig:AKARI_spec}
  \end{center}
\end{figure}

The spectrum of sources C and D taken with {\it AKARI} is well separated 
from that of the other sources (A and B) in \object{II~Zw~096}, 
although they are all in the Np aperture,
because the direction of spectral dispersion is nearly perpendicular to the vector separating
C, D, and A. 
The {\it AKARI} near-infrared spectrum of source C+D is shown in 
Figure~\ref{fig:AKARI_spec}. % and \ref{fig:AKARI_img}.
Strong $3.3\, \micron$ PAH emission, $3.4\, \micron$ emission from aliphatic hydrocarbon grains, 
and Br$\alpha$ emission lines are detected.
The line fluxes are $(2.37 \pm 0.14) \times 10^{-16} \, \mathrm{W\,m^{-2}}$ and 
$(0.50 \pm 0.05) \times 10^{-16} \, \mathrm{W\,m^{-2}}$ for the PAH and 
the Br$\alpha$, respectively, after fitting both with a Gaussian function 
(see Tables~\ref{tbl:spectra} and \ref{tbl:PAHFIT}).
The EQW of the $3.3\, \micron$ PAH is $0.14\, \micron$.
Because the {\it AKARI} spectral resolution between $2.5-5\,\micron$ is only 
$R \sim 120$, the Br$\alpha$ recombination line is not resolved.

% A emission line in the spectrum is fitted by Gaussian function and the line flux is 
% estimated by integrating an interval of the line width, 
% e.g. integration between $3.19-3.37\, \micron$ for the PAH emission line at $3.3\, \micron$ and
% $4.01-4.08\, \micron$ for the Br$\alpha$ emission line at $4.05\, \micron$.

The spectrum of source A (not shown) also exhibits $3.3\, \micron$ PAH and Br$\alpha$ emission.
The $3.3\, \micron$ PAH has a flux of $(1.77 \pm 0.12) \times 10^{-16} \, \mathrm{W\,m^{-2}}$
and an EQW of $0.08\, \micron$. The Br$\alpha$ line flux is 
$(0.33 \pm 0.06) \times 10^{-16} \, \mathrm{W\,m^{-2}}$.

\subsection{Bright Star Clusters In the Remnant} \label{clusters}

The high resolution {\it HST} imaging data show that 
a large number of bright star clusters surround the \object{II~Zw~096} nuclei 
and are present throughout the merging disks (see Figure~\ref{fig:HST_im}).
% a large number of bright star clusters are evident 
% surrounding the buried nuclei and throughout the merging disks (see Figure~\ref{fig:HST_im}).
From the $B$- and $I$-band images, we identify 128 clusters with apparent magnitudes of 
$18.3 \, \mathrm{mag} \leq m_B \leq 26.7 \, \mathrm{mag}$ (Vega) and 
$17.9 \, \mathrm{mag} \leq m_I \leq 25.7 \, \mathrm{mag}$ (Vega) 
with 80\% completeness limits of 26.0~mag and 25.5~mag, respectively.
Of these, 97 and 88 are detected in the FUV and $H$-band images, respectively.
For all clusters, we use a $0.1\arcsec$ radius circular aperture
for the {\it HST} photometric measurements.
We estimate that the clusters contribute $\simeq 14$\% of 
the total FUV luminosity ($\approx 8 \times 10^{36} \, \mathrm{W}$) 
in \object{II~Zw~096} (uncorrected for reddening).

% The single aperture size of $0.1\arcsec$ in radius and the corresponding aperture corrections of 
% $1.46$ (FUV), $2.81$ ($B$-band), $2.30$ (I-band), and $2.48$ ($H$-band) are used for their photometries of the clusters.
% %The photometry results of the clusters are shown in Table~\ref{tbl:clusters}.
% We estimate a ratio of the total emission of clusters to the total 
% emission of the entire galaxy in FUV as approximately 25\% ($L_{FUV} \approx 1.8 \times 10^{36} \, \mathrm{W}$).

\begin{figure}
  \begin{center}
    \includegraphics[scale=1.0, trim=20 0 70 0]{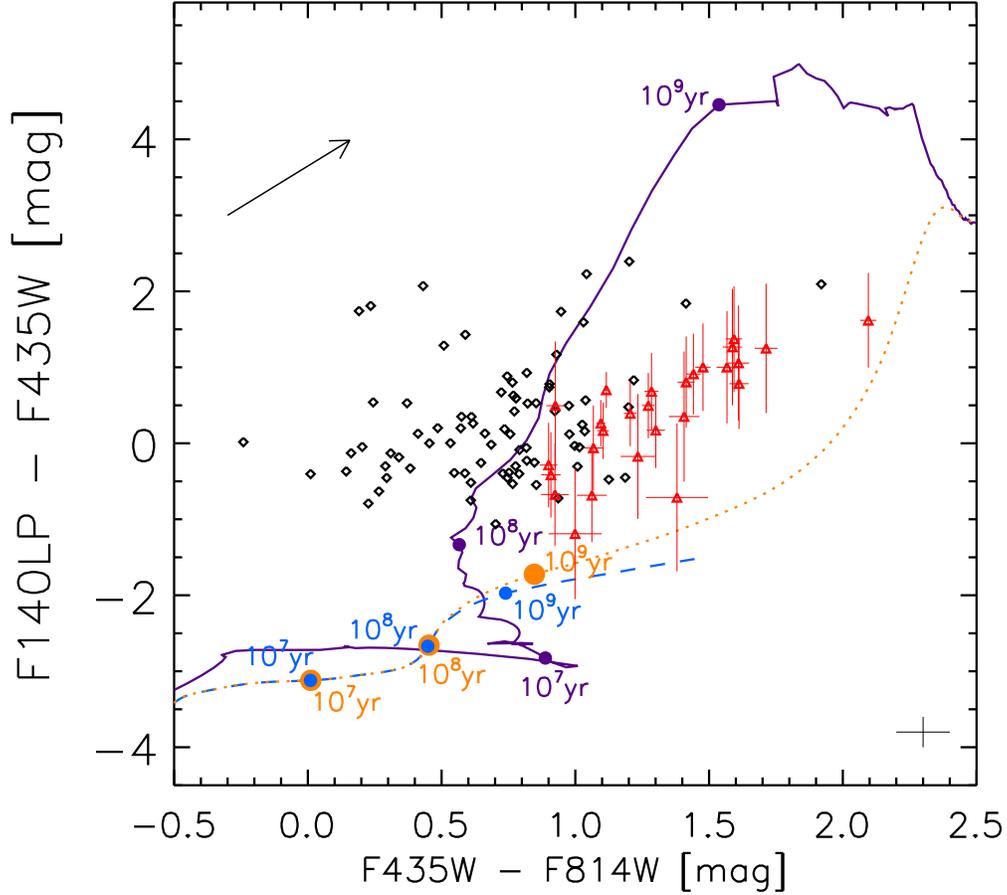}
    \caption{ Optical-UV color-color diagram of the young star clusters (black diamonds) 
      and diffuse emission regions (red triangles) in II~Zw~096.
      Only clusters whose uncertainties are  
      F140LP-F435W~$< 0.2$~mag and F435W-F814W~$< 0.1$~mag are plotted.
      The maximum uncertainty is shown at the lower right as an error bar. 
      The solid purple, the orange dotted and the blue dashed lines are 
      the models that correspond to instantaneous starburst, 
      exponentially declining (SFR=1$M_\sun \tau^{-1} e^{(-t/\tau)}$ for 
      $\tau=1$~Gyr, $0 \leq t \leq 20$~Gyr) and
      continuous star formation age tracks, respectively, with Salpeter IMF 
      and Padova 1994 track \citep{BC03}. The ages are labeled in years.
      The arrow indicates the $A_{V}=1$~mag reddening vector \citep{Calz94}.
      See Figure~\ref{fig:CCD_pos} to identify the positions of the clusters and the diffuse emission regions.}
    \label{fig:CCD}
  \end{center}
\end{figure}

In Figure~\ref{fig:CCD}, we present an FUV-optical (F140LP-F435W vs. F435W-F814W) color-color
diagram of the clusters. We additionally show the colors of a number of diffuse emission
regions throughout the merger system.
For the diffuse emission regions, we use $0.5\arcsec-0.8\arcsec$ radius circular apertures, 
depending on the location in the FUV image to avoid crowded regions.

% Age #1 (All same age)
%-----------------------
Two potential explanations for the observed color distribution are considered.
First, if all the clusters are coeval, 
then %they all formed at the same time during the merger and 
their different colors result from variable extinction.
In other words, those clusters or diffuse emission regions which have colors of
F435W-F814W~$\gtrsim\,1.0$~mag or F140LP-F435W~$\gtrsim\,0.0$~mag
are heavily reddened ($A_V\,\gtrsim\,3$~mag).
When we de-redden an instantaneous starburst model of \cite{BC03},
all of the clusters and the diffuse regions are younger than $\sim$ 5~Myr.
On the other hand, when we pick an exponentially declining
(SFR=1$M_\sun \tau^{-1} e^{(-t/\tau)}$ for $\tau=1$~Gyr, $0 \leq t \leq 20$~Gyr)
or a continuous star formation model \citep{BC03}, 
they have ages between approximately $1-100$~Myr. 
However, we note that the points of cluster colors are
not aligned well with the extinction vector. In particular, 
there is a significant spread in $B-I$ (F435W-F814W) colors, which is not consistent 
with a coeval population with similar intrinsic colors and variable reddening.

% Age #2 (Two populations)
%-----------------------
A more likely explanation is that the colors of the clusters imply at least two populations:
one with F435W-F814W~$< 0.5$~mag and a redder group with F435W-F814W~$\geq 1.0$~mag.
The extinction-corrected ages of the two systems are $1-5$~Myr and $20-200$~Myr 
for the instantaneous starburst model \citep{BC03}, respectively.
We assume that both have colors consistent with reddening by $1-2$ magnitudes of visual extinction ($A_V=1-2$~mag).
However, many clusters have 0.5\,mag~$\leq$~F435W-F814W~$<$~1.0\,mag, 
implying ages of $200-500$~Myr old, if they are un-reddened.
If these clusters sit behind $1-2$~mag of visual extinction, then they could be as young as 5~Myr old.
If we take into account an exponentially declining ($\mathrm{SFR}=1M_\sun \tau^{-1} e^{(-t/\tau)}$ for
$\tau=1$~Gyr, $0 \leq t \leq 20$~Gyr), or a continuous star formation history \citep{BC03},
then the ages of the clusters will be between approximately 1~Myr and 1~Gyr with $A_V=1-3$~mag.
With these models, there are a larger number of older clusters than in the instantaneous starburst model.
In particular the clusters that have colors of both of F435W-F814W~$\gtrsim 1.0$~mag and
F140LP-F814W~$\sim -1.0$~mag, are about $10$ times older.

% Age; vs. Antennae and (U)LIRGs
%-----------------------
In the \object{Antennae Galaxies}, the merger with the most extensively studied system of star clusters,
\cite{Whit99} find three populations of star clusters with ages of
$\sim$10~Myr, $\sim$100~Myr, and $\sim$500~Myr. In the overlap region, the clusters appear to be even younger,
possibly less than 5~Myr old based on H$\alpha$ imaging. The youngest clusters in \object{II~Zw~096}
therefore, appear to have similar ages as the youngest clusters in the \object{Antennae}.
In more luminous galaxies (LIRGs and ULIRGs), \cite{Sura98} and \cite{Scov00} find cluster ages
between $5-300$~Myr, consistent with the spread in ages we find for \object{II~Zw~096}. In \object{Arp~220}, 
\cite{Wils06} similarly find two populations of clusters with ages of $<10$~Myr and $70-500$~Myr.

% Diffuse emission
%-----------------------
The regions of diffuse emission in \object{II~Zw~096} are typically 
$0.5-1$~mag redder in F435W-F814W than the star clusters.
%and $1-2$ mag redder in F140LP-F435W than the star clusters.
If we assume that the emission of the diffuse regions comes from 
a mixture of young stars ($\sim 1$~Myr) and old stars ($\sim 1$~Gyr), 
then the average F140LP-F435W color of the diffuse regions 
can be explained by saying that they contain 
a small fraction ($\sim5-10$\%) of young stars.
%approximately 6\% younger stars. 
On the other hand, the F435W-F814W color of the diffuse emissions suggests
that nearly 17\% of the emission comes from young stars,
although the diffuse emissions have a large spread in this color.

\begin{figure}
  \begin{center}
    \includegraphics[scale=1.0, trim=0 0 110 0]{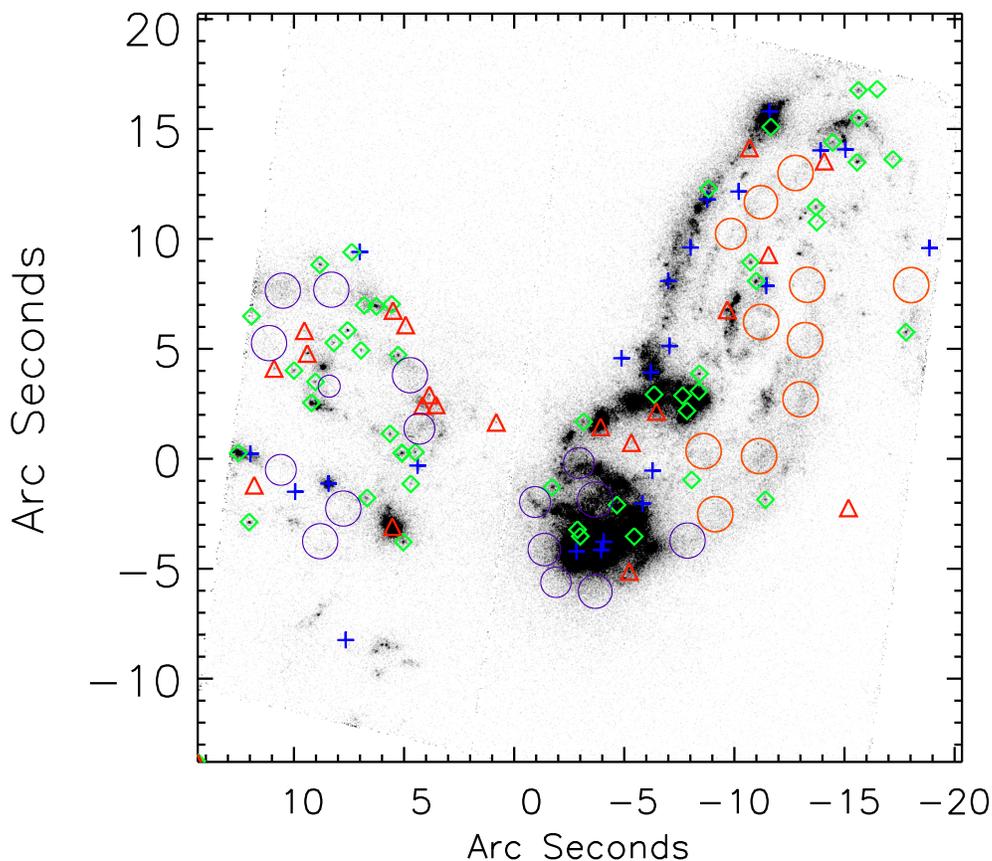}
    \caption{ The {\it HST} ACS SBC F140LP (far-UV) image with the positions of 
      the clusters and diffuse emissions (see section~\ref{clusters}). 
      Blue crosses are the clusters whose F435W-F814W~$< 0.5$~mag implying 
      the ages are $1-5$~Myr, red triangles are those which have F435W-F814W~$\geq 1.0$~mag implying 
      the ages are $20-200$~Myr, and green diamonds have colors between 
      0.5\,mag~$\leq$~F435W-F814W~$<$~1.0\,mag. The purple and orange circles indicate the diffuse emission and 
      correspond to F435W-F814W~$< 1.4$~mag and F435W-F814W~$\geq 1.4$~mag, respectively. 
      The size of the circles corresponds to the aperture size of the photometry measurements. 
      Also see Figure~\ref{fig:CCD}.}
    \label{fig:CCD_pos}
  \end{center}
\end{figure}

% Distribution
%-----------------------
We show the location of the clusters and the diffuse emission on the ACS F140LP 
image in Figure ~\ref{fig:CCD_pos}. 
There is no apparent correlation of the cluster colors (ages) with position,
except for the bluest clusters which are prominent along the spiral arm to the northwest.
%The clusters of various ages are spread uniformly in the system.
On the other hand, the diffuse emission regions with F435W-F814W~$< 1.4$~mag are concentrated in the east 
while those with F435W-F814W~$ > 1.4$~mag are in the west of \object{II~Zw~096}.

The F435W-F160W or F814W-F160W colors of the clusters do not show 
an obvious gradient or pattern with position. 
However, the clusters to the east side have less blue F814W-F160W colors on average than the clusters to the west.

\begin{figure}
  \begin{center}
    \includegraphics[scale=1.0, trim=20 0 70 0]{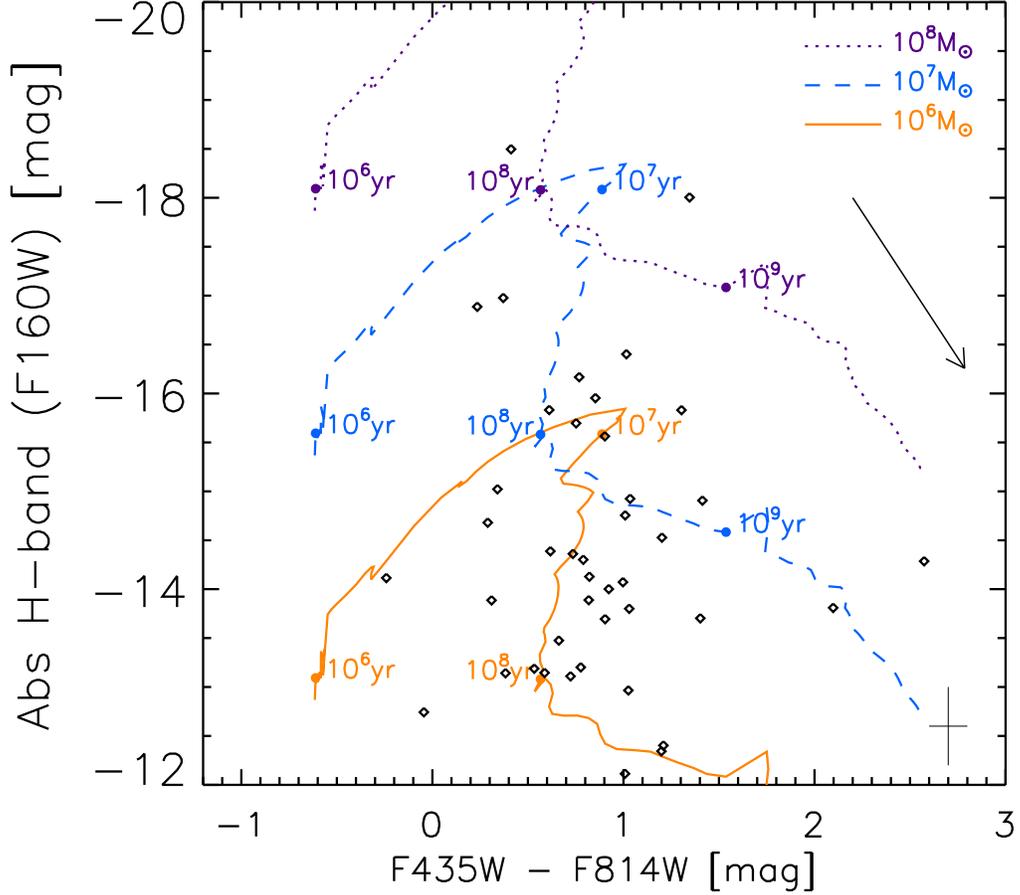}
    \caption{ Optical-Near-infrared color-magnitude diagram of 
      the young star clusters in II~Zw~096. Only the clusters
      which have uncertainties of F160W~$< 0.4$~mag and F435W-F814W~$< 0.1$~mag
      are plotted in the figure. 
      The error bar at the lower right indicates the maximum uncertainty. 
      The dotted, dashed and solid lines 
      % (purple, blue, and orange in the online version) 
      are the instantaneous star burst aging tracks
      for clusters with $ 10^8, 10^7$ and $10^6 \, M_\sun $, respectively. 
      The arrow is the $A_{V}=1$~mag reddening vector.}
    \label{fig:CMD}
  \end{center}
\end{figure}

% Mass of the clusters
%-----------------------
The NICMOS $H$-band data can be used to compute the masses of the clusters.
In Figure~\ref{fig:CMD}, we show an F435W-F814W vs. absolute $H$-band color-magnitude diagram 
for the clusters.  Assuming the clusters have $0-2$~mag of visual extinction, as in the color-color diagram 
(see Figure~\ref{fig:CCD}), the $H$-band data suggest that most of the clusters have masses of $10^{6}-10^{8} \, M_\sun$.

% Mass; vs. Antennae
%-----------------------
If we compare the masses of the star clusters in \object{II~Zw~096} with those in 
the \object{Antennae Galaxies}, the former appear to be $\sim100$ times larger \citep{Zhan99}.
However, we note that the resolution of the HST data is about 30~pc for 
\object{II~Zw~096} as opposed to only 5~pc at the distance of the \object{Antennae}, and 
this might account for the apparent differences in
the cluster masses between the two galaxies.
% However, we note that the clusters in \object{II~Zw~096} 
% may not be resolved and thus this cause an overestimation of luminosity or mass
% as contrasted to the \object{Antennae}.
% The beamsize of {\it HST}/ACS, which we used for observing \object{II~Zw~096}, covers $\sim30$~pc.
% On the other hand, \cite{Whit99} and \cite{Zhan99} used the {\it HST} Wide Field Planetary Camera 2,
% having the beamsize of $\sim5$~pc.
% The bright clusters in \object{II~Zw~096} may be composed of multiple clusters 
% because of the limited resolution.
% Mass; vs. (U)LIRGs
%------------------
For comparison, in (U)LIRGs, the cluster masses derived by \cite{Sura98} and \cite{Scov00} 
lie between $10^5-10^7\,M_\sun$, with the youngest clusters dominating the high mass end.
This is also true in \object{II~Zw~096}, where the youngest clusters 
(those with F435W-F814W~$< 0.5$~mag and $A_V=1-2$~mag) appear to have masses above $10^7\,M_\sun$.

% luminosity function
%-----------------------
In order to compare the distribution of the luminosity of the clusters, 
we derive the $B$-band luminosity function (LF) and fit this with
a power-law, $d\phi/dL_B \propto L_B^{\alpha}$. 
In the range of $-14.7 \,\mathrm{mag} \le M_B \le -10.7 \,\mathrm{mag}$ 
with the detection limit of $-9.9$~mag, we obtain $\alpha=-1.1\pm0.3$.
To avoid confusion due to limited spatial resolution, we chose 
for comparison merging infrared luminous galaxies at similar distances to \object{II~Zw~096}.
The LF of the clusters in \object{II~Zw~096} is found to have a flatter slope ($\alpha=-1.1\pm0.3$)
than that of \object{Mrk~231} ($\alpha=-1.80\pm0.2$) at $z=0.042$ or 
\object{Mrk~463} ($\alpha=-1.63\pm0.11$) at $z=0.050$, which are two nearby ULIRGs studied by \cite{Sura98}.
This suggests that there are more luminous clusters in \object{II~Zw~096} than \object{Mrk~231} and \object{Mrk~463}.
In the \object{Antennae Galaxies} at $z=0.006$, the $V$-band LF of the clusters
have much steeper slopes, $\alpha=-2.6\pm0.2$ and $\alpha=-1.7\pm0.2$
in the ranges of $M_V \lesssim -10.4$~mag and $-10.4$\,mag~$< M_V <$~$-8.0$\,mag, respectively \citep{Whit99}.

\section{Discussion}\label{sec:dis}

We have presented the first mid- and far-infrared images and spectra of \object{II~Zw~096}, 
which show that the extremely red source first identified in the near-infrared 
by \cite{Gold97}, is a heavily obscured starburst that dominates the energetics of
the entire merging system.

\subsection{The Nature of Source C+D}

% AGN/SB DIAGNOSTIC
%-------------------

The IRS spectra of source C+D show the presence of PAH emission, and no evidence
of [\ion{Ne}{5}] or [\ion{O}{4}].  The upper limits on the [\ion{Ne}{5}]/[\ion{Ne}{2}] and 
[\ion{O}{4}]/[\ion{Ne}{2}] emission line flux ratios are 0.04 and 0.11, respectively. 
These line ratios imply a starburst dominated system.
The EQW of the $6.2\,\micron$ PAH feature is $0.26\,\micron$, 
which is only about 1/2 that seen in pure starburst nuclei \citep{Bran06}.  This might indicate
an excess of hot dust in source C+D compared to the nuclei of most starburst galaxies, 
%either due to a more compact source (i.e. a possible obscured galaxy nucleus),
%or a deficiency of PAH emission, or both. 
or a deficiency of PAH emitters.

If we adopt a total far-infrared flux $F_{IR} = 8.48 \times 10^{-13}$ W m$^{-2}$ for source D, and 
a total PAH flux $F_{PAH_{tot}} = 1.44 \times 10^{-14}$ W m$^{-2}$ including all the PAH features (Table~\ref{tbl:PAHFIT}), 
we derive a ratio of $F_{PAH_{tot}} / F_{IR} \, \sim 0.02$. 
This is at the low end for pure starburst galaxies, or about 1/2 the median value \citep{Mars07}.
%$F_{PAH_{tot}}/F_{IR}=1.44e-14/8.48e-13=0.02$

The high [\ion{Ne}{3}]/[\ion{Ne}{2}] line flux ratio of $\sim 1.0$ is at the high-end of that seen in 
starburst galaxies \citep{Thor00,Madd06,Beir08}.
There are two possibilities that can explain this high ratio - either a low metallicity, 
or an unusually hard radiation field, possibly produced by a large number of high mass stars.
According to \cite{Rupk08}, the SE source (source A in this article) has a metallicity of
$0.4 Z_\sun \leq Z \leq 1.8 Z_\sun $, or a mean oxygen abundance $\langle 12+log(\mathrm{O}/\mathrm{H}) \rangle = 8.6$.
% If source C+D has a low metallicity ($Z = 0.2 Z_\sun$, 
% and an ionization parameter $\log{U} = -2.3$), similar to source A,
% then the age of the starburst is $\sim$ 6~Myr (see Figure~10 in \cite{Thor00}).
% On the other hand, 
If we assume solar metallicity for the gas, the implied age is about 3~Myr, and
an upper mass cut off of $50-100\,M_\sun$ is required to match
the observed [\ion{Ne}{3}]/[\ion{Ne}{2}] ratio \citep{Thor00}.

% Another possibility for the weakness of the PAH emission is that 
% high energy photons destroyed the PAH molecules.
% The hardness of the radiation can measured from the [\ion{Ne}{3}]/[\ion{Ne}{2}] ratio and we derived
% $1.0 \pm 0.04$. This value is larger than typical starburst galaxies
% such as \object{M~82} or \object{Arp220}, implying that source D has a ``hard'' radiation field.
% High energetic photons of this hard radiation may have enough energy to break 
% the PAH molecules resulting in the weak PAH emission.

% AKARI spectrum
The strong $3.3\, \micron$ PAH emission feature 
%($(2.37 \pm 0.14) \times 10^{-16} \, \mathrm{W\,m^{-2}}$ and the EQW of $0.14\,\micron$) 
seen in the {\it AKARI} spectrum also supports a powerful starburst in source D \citep{Moor86, Iman08}.
In addition, the typical indicators of a buried AGN, 
e.g. a rising (power law) continuum \citep{Armu07, Veil09}
or $3.4\, \micron$ absorption by aliphatic hydrocarbon grains \citep{Iman06} are not detected.
%and $4.6\, \micron$ CO gas absorption features \citep{Spoo04} 

% X-ray interpolation
The {\it Chandra} data for source C+D is also consistent with a buried starburst, 
although it is slightly weak in X-rays for its far-infrared luminosity.
Starburst galaxies show a correlation between their infrared and X-ray luminosities \citep{Rana03}.
To convert from $2-7$ to $2-10$~keV, we have extrapolated the spectral model 
fit to the data from 7 to 10~keV as in \cite{Iwas09}.
Together, source C+D lie below the starburst correlation, but within the 2-$\sigma$ envelope of the distribution.
In some (U)LIRG nuclei of the GOALS sample, the low ratio of the hard X-ray emission to 
the far-infrared emission has been taken as evidence
for highly obscured AGN \citep{Iwas09}, although this argument hinges on the detection of an Fe~K
emission line. Given the low number of counts in the spectrum of source C+D, an accurate description of the X-Ray
properties will require significantly deeper integrations.
The other X-ray luminous source, source A, lies on the X-ray selected 
starburst correlation line of \cite{Rana03}, suggesting this source is powered by starburst. 

% The {\it Chandra} data for source D is also consistent with a buried starburst, 
% albeit one that is slightly weak in X-rays for its far-infrared luminosity.
% Starburst galaxies show a correlation between their infrared and X-ray luminosities \citep{Rana03}.
% If we assume that the relative emission of sources C and D are 1:1 in the $2-10$~keV X-ray range, 
% and 1:10 in the far-infrared, then source D, and the combination of sources C and D 
% lie below the starburst correlation, but within or slightly lower the $3-\sigma$ scatter.
% To convert from $2-7$ to $2-10$~keV we have extrapolated the spectral model 
% fit to the data from 7 to 10~keV as in \cite{Iwas09}.
% This suggests that a highly obscured population
% of binaries, or even a buried AGN, although the low X-ray counts in \object{II~Zw~096} may be
% insufficient to provide a strong case for an AGN. A much deeper X-ray image is required
% to sort out these options.
% Source C is consistent with the correlation.
% The other X-ray luminous source, source A, lies on the X-ray selected 
% starburst correlation line of \cite{Rana03}, suggesting this source is powered by starburst. 
% %Note that the correction for absorption would not make a large impact 
% %when the absorbing column density is the order of $10^{23} \,\mathrm{cm^{-2}}$.

% SFR
%-----------
From the infrared luminosity of $ L_{IR} = 6.87 \times 10^{11} \, L_\sun$,  
the star formation rate of source D is estimated to be $120\, M_{\sun}  \, \mathrm{ yr^{-1}}$
\citep{Kenn98}. 
This is approximately a factor of five above that estimated by \cite{Gold97} based on the 
near-infrared data alone.
% and using starburst99 models \citep{LH95}.
Assuming this far-infrared emission originates from a region having a size comparable
to that of the projected FWHM of the MIPS $24\, \micron$ beam ($6\arcsec$), 
we estimate a starburst luminosity density of
$ \sim 4.0 \times 10^{10} \, L_\odot \, \mathrm{kpc^{-2}}$, or a star formation rate density of  
$ \sim 6.9 \, M_\sun \, \mathrm{ yr^{-1} \, kpc^{-2}}$. 
If instead, we take the size of source D estimated from the NICMOS image ($\sim 220$~pc radius), 
we obtain values which are about two orders of magnitude larger, namely 
a luminosity density of $ \sim 4.5 \times 10^{12} \, L_\odot \, \mathrm{kpc^{-2}}$ and
a SFR density of $ \sim 780 \, M_\sun \, \mathrm{ yr^{-1} \, kpc^{-2}}$.
Although large, this value is still below the peak values estimated for clusters in starburst galaxies
\cite{Meur97} which are an order of magnitude higher.  Similarly, our derived
value for source D is well below most of the ULIRG starburst nuclei observed by
\cite{Soif00}, except for \object{IRAS~17208-0014}.

%On the other hand, if SFR is estimated by the flux of 6.2 and 11.2 $\mu$m PAH lines 
%\citep{Farr07}, the result is $\sim 32\, M_\sun \, \mathrm{ yr^{-1}}$.
%\cite{Farr07} measured the flux 
%Although the scatter of this estimation is large, about 30-50\%, 
%we can measure the PAH emissions of the source directly without an assumption.

% AKARI Br-alpha SFR
The Br$\alpha$ line detected with {\it AKARI} provides a direct measure of the star formation rate.
We assume Case B for hydrogen recombination, effective temperature of $\sim 10^4$~K, 
density of $\sim 100 \, \mathrm{cm^{-3}}$. % and extinction of $A_V = 0$~mag.
We then estimate a star-formation rate of approximately $45 \, M_{\sun} \, \mathrm{yr^{-1}}$ 
uncorrected for extinction \citep{Kenn98}.
%This is likely to be a lower limit on the SFR because the  Br$\alpha$ has not been corrected for extinction.
Although we do not have an estimate of the extinction from the {\it AKARI} spectra, if we apply the 
line of sight value estimated from the depth of the silicate feature 
in the IRS data (A$_{V} = 19$ mag; \cite{Smit07} eq.(4)), 
the implied SFR would be about $70 \, M_{\sun} \,\mathrm{yr^{-1}}$.

% RADIO/IR RATIO 'q' and SFR
%-----------
Existing radio observations of \object{II~Zw~096} also suggest that source D is a site of intense star formation.
At 1.425~GHz, \cite{Cond96} measure an integrated flux density of 26.6~mJy 
using the Very Large Array (beamsize of $6\arcsec$).
The J2000 position of the radio peak is 
%R.A.=20:57:24.3, Dec=+17:07:38.8, 
R.A.=20$\mathrm{^h}$57$\mathrm{^m}$24$\mathrm{^s}$.3, Dec=+17$\arcdeg$07$\arcmin$38~8$\arcsec$,
%or approximately 0.3 arcsec to the south of source D, 
well within the positional uncertainties of the position of source D in the MIPS data.
%This corresponds to a SFR of $49 \pm 15 \, M_\sun \, \mathrm{ yr^{-1} }$. 
With the new {\it Spitzer} imaging data, we can, for the first time, estimate the infrared to radio ratio
$q$ \citep{Helo85} for source D as $q=2.93$.
This value is within the $\sim1$-$\sigma$ (0.26 dex) scatter of the average $q$ ($\langle q \rangle =2.64$)
among star forming galaxies \citep{Bell03}, and is thus
consistent with our mid-infrared spectra and X-ray data implying starburst activity in source D.

% Mass of source D
%-----------------------
%Source D is invisible in the $B$-band image, so we cannot use Figure~\ref{fig:CMD} 
%to estimate its age. Alternatively, we use 
The $H$-band magnitude ($-19.5$~mag) and assumed age, extinction, and metallicity 
are used to derive the mass of source D by the model of \cite{BC03}.
\cite{Gold97} estimate the age of the \object{II~Zw~096} starburst 
to be roughly $5-7$~Myr from the CO index and Br$\gamma$ equivalent width, 
using the models of \cite{LH95}.
Assuming an extinction of 3.3 mag in $H$-band 
($A_{V} \geq 19 ~\mathrm{mag}$ - see section~\ref{spectra}) and solar metallicity, 
we estimate the mass of source D to be approximately $1-4 \times 10^{9} \, M_\sun$.

% Merger
%-----------
% Although kinematic data are needed to reveal the number of galaxy
% components in the system \object{II~Zw~096}, it seems that 
% it is comprised of two or three merging galaxies.
% % are two prospects: two or three galaxies merger.
% Source B is clearly a spiral disk with the Southern part of the disk being disturbed.
% The very luminous infrared and X-ray emission from source D,
% as well as the highly disturbed morphology of \object{II~Zw~096}, 
% suggest that source D might be the putative starburst nucleus of a third galaxy.
% The high obscuration, even in $H$-band, make it difficult to determine to 
% which galaxy the A and D components belong and what is the exact geometry of 
% the various components in this complex source \object{II~Zw~096}.
% If the system is comprised of only two merging galaxies, then the large amount of 
% diffuse emission at $\sim10\arcsec$ east of sources C and D in the $B$- and $I$-band 
% images may come from a part of the collapsed disk of source A. 
% The latter scenario is more likely that source A merges with source B and then this causes
% the offnuclear objects, sources C and D in this system.
% %Then either source A or D has its component of an underlying galaxy, 
% which is not shown because of high obscuration even in $H$-band. If this system is two galaxy collision, 
% a large amount of  diffuse emission at $\sim10\arcsec$ east of sources C and D in 
% the $B$- and $I$-band images can be implied to be a part of collapsed disk of source A.

\subsection{The Comparison with Other Merger-Induced Extranuclear Starbursts}

%MEGER-INDUCED OFF-NUCLEAR SB
%---------------------------------

Source D is a powerful starburst not associated with 
the primary nuclei of the \object{II~Zw~096} system (sources A and B).
It could be a starburst in the disturbed disk of source A, or even the nucleus
of a third galaxy - one which is nearly completely hidden behind A and B.
While large scale mapping and modeling of the gas and stellar dynamics of \object{II~Zw~096}
will be needed to fully understand the true nature of the system, we favor the 
former explanation for two reasons.  First, simulations have shown (e.g. \cite{Barn04}) that shock-induced 
starbursts can occur both in and outside the nuclei of merging galaxies.  Second, we have a 
number of examples of powerful, extranuclear starbursts in the local Universe.
Below we compare the properties of source D to the extra-nuclear starbursts in 
two well-studied merging galaxies \object{NGC~4038/9} (\object{the Antennae Galaxies}) and \object{Arp~299}.

%Simulations of shock-induced star formation \citep{Barn04} claim that 
%when galaxies are in a relatively early merging stage, 
%there is vigorous star-formation not only in the nuclei of the merging galaxies 
%but also throughout the tidal tails. 
%Here, we may have found direct evidence for this in source D 
%which appears to be an extremely luminous merger-induced extranuclear starburst. 
%Below we compare source D with other similar extranuclear objects.  

% COMPARISON WITH the ANTENNAE
%--------------------------
In \object{the Antennae Galaxies}, nearly $1/2$
of the total $15\,\micron$ emission (about $5\times 10^{10} L_\sun$) comes from 
the overlap region ($5\times3$~kpc) between the merging spiral galaxies, 
with about $15\%$ coming from a single starburst knot \citep{Mira98}. 
Recent mid-infrared spectral mapping observations with the IRS on {\it Spitzer} have resolved
this off-nuclear starburst region into a number of clumps in the mid-infrared 
(peaks 1, 2, 3, and 5 of \cite{Bran09}).
% LIR
%------
% One of the most remarkable results in the work presented here, is that source D dominates 
% the $\sim 80\%$ of the total $L_{IR}$ of the entire \object{II~Zw~096} system.
These clumps (or clusters) have infrared luminosities of 
$3.86 \times 10^{9} L_{\sun} < L_{IR} < 1.14 \times 10^{10} L_{\sun} $,
corresponding to $\sim 5-16$~\% of infrared radiation coming from the system. %\citep{Bran09}.
% SFR
%------
\cite{Bran09} estimate that the SFRs in these clusters range between 
$0.66-1.97 \, M_{\sun}\mathrm{yr^{-1}}$.

% LIRD
%------
% To make a direct comparison with \object{II~Zw~096}, 
% we assume the radius of the \object{Antennae} cluster $R\sim500\,\mathrm{pc}$ 
% and obtain a luminosity density of $(0.5-1.5) \times 10^{10} \, L_{\sun}\,\mathrm{kpc^{-2}}$.
% This value is smaller than the luminosity density of source D in \object{II~Zw~096} 
% even when the beamsize of the MIPS $24\,\micron$ is used for its size.
% These results suggest \object{II~Zw~096} is an object similar to
% a scaled-up version of the \object{Antennae Galaxies}.

To make a direct comparison with \object{II~Zw~096}, 
we pick the most luminous infrared peak in the \object{Antennae} (peak~1 in \cite{Bran09})
which has $L_{IR} = 1.14 \times 10^{10} L_{\sun}$ and assume the radius of this 
cluster to be $R\sim48\,\mathrm{pc}$ (the size estimated by \cite{Whit95}).
We then estimate a luminosity density of  $1.6 \times 10^{12} L_{\sun}\,\mathrm{kpc^{-2}}$.
This value is about three times smaller than the upper limit of the luminosity density of source D 
(assuming a size of 220 pc radius - the size of D in the NICMOS image).  While the 
luminosity density of peak~1 is extremely large, it only contributes $\sim15\%$ of the
total infrared emission of the entire system, in contrast to
source D in \object{II~Zw~096} which dominates the total infrared output of the system.
%luminosity density more than an order of magnitude larger than the brightest knot in the overlap region. 
% These results suggest \object{II~Zw~096} is an object similar to
% a scaled-up version of the \object{Antennae Galaxies}.

% Extinction
%------
In the \object{Antennae Galaxies}, the silicate optical depths at $9.7\,\micron$ of the star clusters
measured by \cite{Bran09} are $\tau_{9.7} = 0.13-1.03$ ($A_V=2.4-19$~mag). 
Although one of the clusters in the \object{Antennae}, peak~5, has an extinction
comparable to that measured in source D in \object{II~Zw~096} ($\tau_{9.7}\sim1.03$),
its $L_{IR}$ is the smallest in all of the clusters in the overlap region of the \object{Antennae}. 
This is unlike source D which has both high extinction and high $L_{IR}$.
Peak~1 itself, has a $\tau_{9.7} = 0.19$, much lower than our estimated
value for source D, even though the estimated age of peak~1 ($2-4$ Myr)
is slightly less than that derived for source D ($5-7$ Myr - \cite{Gold97}).

% Age, H2 & [NeIII]/[NeII]
%------
%The ages of the clusters in the overlap region of the \object{Antennae} are between $2-4$~Myr, 
%slightly lower than the age estimated for source D ($5-7$ Myr), estimated by \cite{Gold97}. 
% Interestingly, the clusters in the non-overlap region (Peaks 4 and 6) in the \object{Antennae} 
% have ages of 6 and 7~Myr, suggesting they are older than those in the overlap region 
% but similar age to source D \cite{Gold97}.
%The lower ratios of [\ion{Ne}{3}]/[\ion{Ne}{2}] ($0.6-0.75$) and the lower $\mathrm{H_2}$ temperature 
%of the clusters in the \object{Antennae} compared to those in \object{II~Zw~096}, 
%may imply that source D has been forming stars for a more extended period than the \object{Antennae}.

% COMPARISON WITH the ARP299
%--------------------------
\object{Arp~299} (\object{IC~694} + \object{NGC~3690}) has
substantial extranuclear infrared emission like the \object{Antennae} but 
an infrared luminosity of $\sim 6 \times 10^{11} \, L_{\sun}$, making this system a LIRG. 
In particular, regions C and C' in the overlap region between 
\object{IC~694} and \object{NGC~3690} have all the signatures of a merger-induced 
starburst (e.g. \cite{Alon00, Char02, Alon09}).
% LIR
%------
Region C+C' have an infrared luminosity ($4.4 \times 10^{10} \, L_\odot$; \cite{Char02}),
approximately 6.4\% of the $L_{IR}$ of source D in \object{II~Zw~096}.
The fraction of the infrared emission in regions C and C' to 
that in the \object{Arp~299} system is only $\sim 10\%$ \citep{Char02}, 
much less than the fractional infrared luminosity in source D (80\%).
The infrared luminosity densities of regions C and C' are estimated to be 
$ \sim 1.0 \times 10^{11} \, L_\odot \, \mathrm{kpc^{-2}}$ and 
$ \sim 7.5 \times 10^{10} \, L_\odot \, \mathrm{kpc^{-2}}$, respectively,
if we assume that their sizes are $R\sim300\,\mathrm{pc}$ and $R\sim250\,\mathrm{pc}$,
corresponding to the size of Pa$\alpha$ emission estimated by \cite{Alon09}, 
and that the infrared luminosity of regions C and C' have a ratio of 2:1. 
These numbers are a factor of $2-3$ larger than what we estimate for source D 
when we assume that the size of the source D is equivalent to the beamsize of the MIPS $24\,\micron$. 
However, if the size of source D is comparable to the NICMOS resolution,
then the luminosity density could be significantly larger than that seen in 
regions C and C' in \object{Arp~299}.
% then the luminosity density could be significantly larger,
% e.g. as indicated by the size of the NICMOS source above, 
% suggesting source D in \object{II~Zw~096} may have luminosity density $40-50$ times 
% greater than regions C and C' in \object{Arp~299}.
% A higher resolution far-infrared image is needed to place more stringent limits on this estimate. 

% Age
%------
Assuming a metallicity of $2Z_{\sun}$, \cite{Alon09} use the IRS spectra of regions C and C' 
and the models of \cite{Snij07} to estimate ages of $4-7$~Myr.
If we apply the same model and metallicity to source D in \object{II~Zw~096},
the measured [\ion{Ne}{3}]/[\ion{Ne}{2}]~$=1.0$ and [\ion{S}{4}]/[\ion{S}{3}]~$=0.33$ line ratios
suggest that source D also has an age of $4-7$ Myr.

\section{Summary and Conclusions}\label{sec:con}

The first analysis of the interacting galaxy \object{II~Zw~96} utilizing mid- and
far-infrared images and mid-infrared spectroscopy leads to the following conclusions:

\begin{enumerate}

\item
The {\it Spitzer} imaging reveals that approximately 80\% of 
the far-infrared emission of the \object{II~Zw~096} comes from a compact, off-nuclear
starburst with 
R.A.=20$\mathrm{^h}$57$\mathrm{^m}$24$\mathrm{^s}$.34, Dec=+17$\arcdeg$07$\arcmin$39~1$\arcsec$ (J2000).
The estimated $8-1000 \, \micron$ luminosity of this source is
$ L_{IR} = 6.87 \times 10^{11} \, L_\sun$. 
The implied star formation rate of this object is about $120 \, M_\sun \, \mathrm{yr^{-1}}$.

\item
{\it HST} NICMOS observations show that the off-nuclear starburst is 
composed of two prominent, red knots, with 
a number of smaller peaks spread over approximately 10 square arcseconds ($\sim 6.6 \, \mathrm{kpc^2}$). 
Most of this emission is not seen in the optical imaging data, either from the ground or with the {\it HST} ACS.
The knot to the southwest of this region, source D, is consistent with the peak in the {\it Spitzer} MIPS 24 and 70 $\mu$m
emission which dominates the far-infrared luminosity of the entire merging galaxy.
Assuming a standard mass-to-light ratio for galaxies, the H-band
luminosity of source D corresponds to a mass of $1-4 \times 10^{9} \, M_\sun$.

\item
The {\it Chandra} X-ray imaging shows that the hardest source in the system
is source C+D, and it lies slightly below the correlation of X-ray and infrared luminosities 
seen for starburst galaxies, although within the 2-$\sigma$ scatter of the relation.

\item
The {\it Spitzer} IRS and {\it AKARI} spectra suggest a starburst-dominated system with 
strong $3.3\mu$m and $6.2\mu$m PAH emission, and no evidence of emission from 
[\ion{Ne}{5}] $14.3\mu$m or [\ion{O}{4}] $25.9\mu$m in the IRS data.  While the 
equivalent width of the $3.3\mu$m PAH feature as measured in the {\it AKARI} spectrum is
typical for starburst galaxies, the equivalent width of the $6.2\mu$m PAH feature as 
measured in the IRS spectrum is about 1/2 the value found in pure starbursts.  This
may reflect an excess of hot dust surrounding source C+D, which is easier to detect in 
the narrow IRS slit.  This is consistent with the [\ion{Ne}{3}]/[\ion{Ne}{2}]
line flux ratio of $\sim1$ indicating a relatively hard radiation field. The $9.7\, \micron$ silicate 
optical depth, measured from the IRS spectra suggests an $A_V \geq 19$~mag 
toward source C+D, indicating a highly buried source.

\item
The ACS $B$ and $I$-band imaging reveals a large number of (super) star clusters in the \object{II~Zw~096}
system, consisting of at least two populations: one $1-5$~Myr and 
one $20-500$~Myr, which may have formed at different times during the merger.
We find no clear association of cluster age with position in the merger.
% There is no preference for clusters of a certain age to be found in one region of the merger. 
Most of the cluster masses are in the range of $10^6 - 10^8 \, M_\sun$.

\end{enumerate}

Spitzer imaging and spectroscopy of \object{II~Zw~096} has revealed a powerful, 
young extranuclear starburst.  This starburst is reminiscent of those
seen in \object{NGC~4038/9} (\object{the Antennae Galaxies}) and \object{Arp~299}, but it is more
luminous (more than an order of magnitude more luminous than the Antennae starburst),
and it is responsible for nearly all the infrared luminosity in 
\object{II~Zw~096} (compared to only $10-15\%$ in \object{the Antennae} or \object{Arp~299}).
Source D in \object{II~Zw~096} is one of the most extreme buried extra-nuclear starbursts yet
discovered in the local Universe.

\acknowledgments

We thank J. Goldader and S. Satyapal for many helpful suggestions.
The authors also wish to thank an anonymous referee for suggestions which improved the manuscript.
The {\it Spitzer} Space Telescope is operated by the Jet Propulsion Laboratory, 
California Institute of Technology, under NASA contract 1407. 
This research has made use of the NASA/IPAC Extragalactic Database (NED) 
and the Infrared Science Archive (IRSA) 
which are operated by the Jet Propulsion Laboratory, California Institute of Technology, 
under contract with the National Aeronautics and Space Administration.
Hanae Inami thanks the {\it Spitzer} Visiting Graduate Student Fellowship (from Sep 2008 to Feb 2009)
and Grant-in-Aid for Japan Society for the Promotion of Science (JSPS) 
Fellows (21-969) for supporting this work.


%% in parentheses, after the keyword, but they will not be verified.

{\it Facilities:} \facility{Spitzer}, \facility{HST}, \facility{CXO}, \facility{Akari}

\bibliography{GOALS}

\clearpage

\clearpage

\clearpage

\clearpage

\clearpage

\clearpage

\begin{table}
  \begin{center}
    \caption{ {\it Spitzer}, {\it Hubble}, {\it Chandra}, and {\it AKARI} observations of II~Zw~096 \label{tbl:obs}}
    \begin{tabular}{|c|cccc|}
      \tableline\tableline
      Observatory & Instrument & Filter or module & Integration time [sec] & Observation date  \\
      \tableline
      \multirow{11}{*}{{\it Spitzer}} & \multirow{4}{*}{IRAC}        & $3.6\,\micron$ & 150  & \multirow{4}{*}{2004 Oct 29}  \\
                                                  &                                          & $4.5\,\micron$ & 150  &                                                 \\
                                                  &                                          & $5.8\,\micron$ & 150  &                                                 \\
                                                  &                                          & $8.0\,\micron$ &   6    &                                                 \\
                                                  \cline{2-5}
                                                  & \multirow{3}{*}{MIPS}         & $24 \,\micron$ & 48.2  & \multirow{3}{*}{2004 Nov 30} \\
                                                  &                                          & $70 \,\micron$ & 37.7  &                                                 \\
                                                  &                                          & $160\,\micron$& 25.2  &                                                 \\
                                        \cline{2-5}
                                        & \multirow{4}{*}{IRS} & SL & $6\times14$\tablenotemark{a}  & \multirow{4}{*}{2006 Nov 12} \\
                                        &                                & LL & $4\times14$\tablenotemark{a}  &                                                \\
                                        &                                & SH & $6\times30$\tablenotemark{a} &                                                \\
                                        &                                & LH & $4\times60$\tablenotemark{a} &                                                \\
      \tableline
      \multirow{4}{*}{{\it HST}}     & ACS/SBC                            & F140LP              & 2528 & 2008 May 1                              \\
                                                  \cline{2-5}
                                                  & \multirow{2}{*}{ACS/WFC} & F435W               & 2520 & \multirow{2}{*}{2006 Apr 15}  \\
                                                  &                                          & F814W               & 1440 &                                                 \\
                                                  \cline{2-5}
                                                  & NICMOS/NIC2                   & F160W               & 2304 & 2008 Jun 1                               \\
      \tableline
      {\it Chandra}                        & ACIS                                  & $0.5-7$~keV    & $14.56 \times 10^3$ & 2007 Sep 10   \\
      \tableline
      {\it AKARI}                            & IRC                                    & NG with Np       & 799   & 2009 May 12 \& 13                  \\
      \tableline
    \end{tabular}
    \tablenotetext{a}{number of cycles $\times$ integration time in each nod position}
  \end{center}
\end{table}

%\clearpage

\begin{table}
  \begin{center}
    \caption{ Fine structure and molecular lines \label{tbl:spectra}}
    \begin{tabular}{ccc}
      \tableline\tableline
      Line & wavelength [$\micron$] & flux [$\times 10^{-16} \mathrm{W\,m^{-2}}$]\\
      \tableline
%      \,PAH \tablenotemark{a}  & 6.2 &  $4.7 \pm 0.3 $               \\
      \, Br$\alpha$            &  4.050 & $0.50 \pm 0.05$ \\
      \, H$_2~S(3)$            &  9.660 & $0.45 \pm 0.05$ \\
      \,[\ion{S}{4}]               & 10.511 & $0.72 \pm 0.04$ \\
      \,H$_2~S(2)$             & 12.279 & $0.29 \pm 0.03$ \\  % $0.293 \pm 0.059$
      \,[\ion{Ne}{2}]            & 12.814 & $2.74 \pm 0.47$ \\  % $2.49 \pm 0.17 $  \\
      \,[\ion{Ne}{5}]            & 14.322 & $< 0.10 $ \tablenotemark{a}  \\
      \,[\ion{Ne}{3}]            & 15.555 & $2.79 \pm 0.03$ \\  % $2.73 \pm 0.07 $  \\
      \,H$_2~S(1)$             & 17.035 & $0.73 \pm 0.05$ \\ % $0.60 \pm 0.20$
      \,[\ion{S}{3}]               & 18.713 & $2.16 \pm 0.22$ (SH) \\   % $1.64 \pm 0.06$
      \,[\ion{S}{3}]               & 18.713 & $3.27 \pm 0.04$ (LH) \\   % $3.04 \pm 0.02$
      \,[\ion{O}{4}]              & 25.890 & $< 0.29$ \tablenotemark{a}   \\  %$< 0.24$ \tablenotemark{a} \\
      \,[\ion{Fe}{2}]             & 25.988 & $0.56 \pm 0.22$ \\ %$0.51 \pm 0.03$   \\
      \,[\ion{S}{3}]               & 33.481 & $4.77 \pm 0.07$ \\ % $4.5 \pm 0.1$  \\
      \,[\ion{Si}{2}]              & 34.815 & $4.23 \pm 0.12$ \\ % $4.8 \pm 0.2$
      \tableline
    \end{tabular}
%    \tablenotetext{a}{ This line is detected by {\it AKARI} from sources C and D. 
 %     The rest of the lines are detected by {\it Spitzer} from source D. }
    \tablenotetext{a}{3-$\sigma$ upper limit}
    \tablecomments{The LH spectra cover not only source C+D but also source A (see text for the detail).}
  \end{center}
\end{table}

\begin{table}
  \begin{center}
    \caption{ PAH features \label{tbl:PAHFIT}}
    \begin{tabular}{ccc}
      \tableline\tableline
      wavelength [$\micron$] & flux [$\times 10^{-16} \mathrm{W\,m^{-2}}$] & EQW [$\micron$] \\
%      & $F_{PAH}/F_{PAH_{tot}}$ \tablenotemark{a} & 
%      mean SB $F_{PAH}/F_{PAH_{tot}}$ \tablenotemark{b} \\
      \tableline
      \, 3.3 \tablenotemark{a}   &   $2.37 \pm 0.14$  & 0.14  \\
      \, 6.2 \tablenotemark{b}   &  $5.57 \pm  0.19$  & 0.26  \\
      \, 11.2 \tablenotemark{b} &  $3.24 \pm  0.32$  & 0.14  \\
%      \cline{2-3}
      \hline
      \, 6.2                                &  $14.11 \pm 0.35$  & 1.30 \\  % & 0.098  & 0.111\\
      \, 7.7 complex                 &  $83.57 \pm 1.14$  & 7.61 \\  % & 0.578 & 0.429 \\
      \, 8.3                                & $4.55 \pm 0.44$  & 0.35  \\ % &  & ... \\
      \, 8.6                                & $12.86 \pm 0.37$  & 0.99  \\ % &  & ... \\
      \, 11.3 complex               & $4.98 \pm 0.15$  & 0.16  \\ % & 0.034 & 0.087 \\
%      \, 12.0                            &  ...                            & ...          & ...       & ... \\
      \, 12.6                             & $3.34 \pm 0.27$  & 0.05  \\ % &  & ... \\
      \, 13.6                             & $0.50 \pm 0.24$  & 0.01  \\ % &  & ...   \\
      \, 14.2                             & $7.08 \pm 0.26$  & 0.09  \\ % &  & ...  \\
      \, 17 complex                 & $13.50 \pm 0.61$  & 0.15 \\  % &  & ...  \\
%      \, $F_{PAH_{tot}}$         & $144.49 \pm $ ...  & ... & ... & \\
      \tableline
    \end{tabular}
%    \tablenotetext{a}{Upper limit (because we have not summed up all PAH fluxes which \cite{Mars07} have done)}
%    \tablenotetext{b}{Referred to table 12 of \cite{Mars07}}
    \tablenotetext{a}{ The {\it AKARI} detection. The others are detected by {\it Spitzer}. 
      The {\it AKARI} spectrum is a mixture of sources C and D but the {\it Spitzer} spectra represent source D.}
    \tablenotetext{b}{Measured using a simple spline fit to the continuum, and integrating the feature flux above the continuum.  Other values are measured using the PAHFIT routine (Smith et al. 2007), which fits Drude profiles to all the PAH features, resulting in systematically larger fluxes and EQW}.
  \end{center}
\end{table}

%\begin{deluxetable}{|c|ccccc|}
\begin{deluxetable}{cccccc}
  \tabletypesize{\scriptsize}
  \rotate
  \tablecaption{{\it Spitzer}, {\it HST}, and {\it Chandra} Photometries of II~Zw~096 \label{tbl:phot}}
  \tablewidth{0pt}
  \tablehead{
    \colhead{ \multirow{2}{*}{Observatory}} & \colhead{ \multirow{2}{*}{Instrument}} & \colhead{ \multirow{2}{*}{Filter or module}} & 
    \colhead{Source A} & \colhead{Source C} & \colhead{Source D} \\
     &  &  & 
    \colhead{20$\mathrm{^h}$57$\mathrm{^m}$24$\mathrm{^s}$.09, +17$\arcdeg$07$\arcmin$35~3$\arcsec$} & 
    \colhead{20$\mathrm{^h}$57$\mathrm{^m}$24$\mathrm{^s}$.47, +17$\arcdeg$07$\arcmin$39~9$\arcsec$} & 
    \colhead{20$\mathrm{^h}$57$\mathrm{^m}$24$\mathrm{^s}$.34, +17$\arcdeg$07$\arcmin$39~1$\arcsec$} 
  }
  \startdata
  \multirow{7}{*}{{\it Spitzer}} & \multirow{4}{*}{IRAC}    & $3.6\,\micron$ & $3.29\pm0.10$ ($2.9\arcsec$) & $2.26\pm0.08$ ($1.4\arcsec$) & $2.85\pm0.08$ ($1.4\arcsec$)   \\
                                 &                          & $4.5\,\micron$ & $2.68\pm0.08$ ($2.9\arcsec$) & $3.33\pm0.08$ ($1.4\arcsec$) & $4.67\pm0.13$ ($1.4\arcsec$)   \\
                                 &                          & $5.8\,\micron$ & $8.43\pm0.18$ ($2.9\arcsec$) & $9.36\pm0.22$ ($1.4\arcsec$) & $12.7\pm0.2$ ($1.4\arcsec$) \\
                                 &                          & $8.0\,\micron$ & $28.1\pm0.5$ ($2.9\arcsec$)  & $27.2\pm0.4$ ($1.4\arcsec$)  & $41.9\pm0.8$ ($1.4\arcsec$) \\
                                 \cline{2-6}
                                 & \multirow{3}{*}{MIPS}    & $24 \,\micron$ & ... & ... & $1.37\pm0.18$ \tablenotemark{a}  \\
                                 &                          & $70 \,\micron$ & ... & ... & $10.2\pm1.2$ \tablenotemark{a}   \\
                                 &                          & $160\,\micron$ & ... & ... & ...  \\
  \tableline
  \multirow{6}{*}{{\it HST}}     & ACS/SBC                  & F140LP  & $(6.59\pm0.30)\times10^{-15}$ ($2.9\arcsec$) & ... & ...  \\
                                 \cline{2-6}
                                 & \multirow{3}{*}{ACS/WFC} & \multirow{2}{*}{F435W} & \multirow{2}{*}{$(3.03\pm0.02)\times10^{-15}$ ($2.9\arcsec$) } & \multirow{2}{*}{...}                         & $(3.43\pm0.34)\times10^{-17}$ ($1.4\arcsec$)  \\
                                 &                          &                        &                                               &                                              & $(4.63\pm0.67)\times10^{-18}$ ($0.25\arcsec$) \\
                                 &                          & \multirow{2}{*}{F814W} & \multirow{2}{*}{$(1.28\pm0.01)\times10^{-15}$ ($2.9\arcsec$) } & $(5.57\pm0.19)\times10^{-17}$ ($1.4\arcsec$)  & $(4.85\pm0.20)\times10^{-17}$ ($1.4\arcsec$)  \\
                                 &                          &                        &                                               & $(4.64\pm0.39)\times10^{-18}$ ($0.25\arcsec$) & $(2.85\pm0.08)\times10^{-17}$ ($0.25\arcsec$) \\
                                 \cline{2-6}
                                 & \multirow{2}{*}{NICMOS/NIC2} & \multirow{2}{*}{F160W} & \multirow{2}{*}{$5.36\pm0.05$ ($2.9\arcsec$)} & $0.85\pm0.02$ ($1.4\arcsec$)  & $0.74\pm0.02$ ($1.4\arcsec$)      \\
                                 &                              &                        &                                               & $0.13\pm0.01$ ($0.25\arcsec$) & $0.24\pm0.01$ ($0.25\arcsec$) \\
  \tableline
  \multirow{2}{*}{{\it Chandra}} & \multirow{2}{*}{ACIS}    & $0.5-2$~keV    & $6.6\times10^{40}$ &   \multicolumn{2}{c}{$1.8\times10^{40}$~(Source C+D)} \\
                                 &                          & $2-7$~keV      & $1\times10^{41}$   & \multicolumn{2}{c}{$1\times10^{41}$~(Source C+D)}     \\
  \enddata
  \tablecomments{The values in the brackets are the circular aperture sizes in radii. The units are mJy for IRAC and NICMOS/NIC2, Jy for MIPS, 
                 $\mathrm{erg\,s^{-1}\,cm^{-2}\,A^{-1}}$ for ACS/SBC and ACS/WFC, and $\mathrm{erg\,s^{-1}}$ for ACIS.}
  \tablenotetext{a}{ Measured by the PSF fitting (see \S\ref{IRemission}). }
\end{deluxetable}

\end{document}